\documentclass{aa}  
\usepackage{graphicx, natbib, amsmath, longtable}
\bibpunct{(}{)}{;}{a}{}{,}
\begin{document}
   \title{Spectroscopic monitoring of the Luminous Blue Variable Westerlund1-243 from 2002 to 2009\thanks{This work is 
     based on observations collected at the European Southern Observatory, 
     Chile, under programme IDs ESO 69.D-0039(A), 071.D-0151(A), 271.D-5045(A), 
     073.D-0025(A \dots C), 075.D-0388(A \dots C), 081.D-0324(A \dots F) and 383.D-0633(A).}}

   \author{B.~W.~Ritchie\inst{1,2} \and J.~S.~Clark\inst{1} \and 
      I.~Negueruela\inst{3} \and F.~Najarro\inst{4}}
   \offprints{B.W.~Ritchie, \email{b.ritchie@open.ac.uk}}

   \institute{
     Department of Physics and Astronomy, The Open University, 
     Walton Hall, Milton Keynes MK7 6AA, United Kingdom
   \and
     IBM United Kingdom Laboratories, Hursley Park, Winchester, 
     SO21 2JN, United Kingdom
   \and 
     Departamento de F\'{\i}sica, Ingenier\'{\i}a de Sistemas y 
     Teor\'{\i}a de la Se\~{n}al, Universidad de Alicante,
     Apdo. 99, 03080 Alicante, Spain
   \and 
     Laboratorio de F\'{\i}sica Estelar y Exoplanetas, Centro de Astrobiolog\'{\i}a, CSIC-INTA,
     Ctra de Torrej\'on a Ajalvir km~4, 28850 Torrej\'on de Ardoz, Spain
   }

   \date{Accepted ??? Received ???}

   \abstract
   {The massive post-Main Sequence star W243 in the galactic starburst cluster 
   Westerlund 1 has undergone a spectral transformation from a B2Ia 
   supergiant devoid of emission features in 1981 to an A-type supergiant with 
   a rich emission-line spectrum by 2002/03. The star was proposed as a 
   Luminous Blue Variable undergoing an eruptive event by Clark \& 
   Negueruela (2004).}
   {We examine the continued evolution of W243 from 2002 to 2009 to understand 
    its evolutionary state, current physical properties and the origin of its 
    peculiar emission line spectrum.}
   {We used VLT/UVES and VLT/FLAMES to obtain high resolution, high 
    signal-to-noise spectra on six epochs in 2003/04 (UVES) and ten
    epochs in 2008/09 (FLAMES). These spectra are used alongside other 
    lower-resolution VLT/FLAMES and NTT/EMMI spectra to follow the evolution of 
    W243 from 2002 to 2009. Non-LTE models are used to determine 
    the physical properties of W243.}
   {W243 displays a complex, time-varying spectrum with emission lines of 
    Hydrogen, Helium and Lyman-$\alpha$ pumped metals, forbidden lines of 
    Nitrogen and Iron, and a large number of absorption lines from neutral 
    and singly-ionized metals. Many lines are complex emission/absorption 
    blends, with significant spectral evolution occurring on timescales of 
    just a few days. The LBV has a temperature of $\sim$8500K  
    (spectral type A3Ia$^{+}$), and displays signs of photospheric pulsations
    and weak episodic mass loss. Nitrogen is highly overabundant, with Carbon
    and Oxygen depleted, indicative of surface CNO-processed material and 
    considerable previous mass-loss, although current time-averaged mass-loss 
    rates are low. The emission-line spectrum forms at large radii, when
    material lost by the LBV in a previous mass-loss event is ionized by an
    unseen hot companion. Monitoring of the near-infrared spectrum 
    suggests that the star has not changed significantly since it finished 
    evolving to the cool state, close to the Humphreys-Davidson limit, in 
    early 2003.}
   {}

   \titlerunning{The LBV Wd~1-243}

  \maketitle
%

\section{Introduction}

Luminous Blue Variables (LBVs; \citealt{hd94}) represent a short-lived 
transitional stage in the life of the most massive stars as they leave 
the main sequence and evolve towards the Wolf-Rayet phase. LBVs are 
characterized by very high rates of mass-loss, dense, slow winds and 
variability in luminosity and temperature that range from short-term 
microvariations of $\sim$0.1 to 0.2 magnitudes to the rare, catastrophic 
\object{$\eta$~Carinae} type eruptions for which the LBV 
class is most well known. In its quiescent state a LBV appears as a very 
luminous ($\sim$10$^6 \text{L}_{\odot}$) B-type supergiant that lies on the 
\object{S~Doradus} instability strip \citep{svk04}, but on timescales of a 
few decades the LBV will move redwards on the Hertzsprung-Russell 
Diagram (HRD), taking on the appearance of an A- or F-type supergiant and brightening 
by $m_{v}$$\sim$1~to~2 due to the change in bolometric correction. More
rarely, LBVs may undergo \textit{giant} eruptions in which bolometric luminosity
is not conserved and substantial mass-loss occurs, with many LBVs surrounded by 
expanding nebule (e.g. \citealt{w03}) that consist of material ejected from the star during 
earlier mass-loss events. Finally, there is growing evidence that LBVs may be the 
progenitors of some type~IIn supernovae (e.g. \object{SN2006gy}, 
\citealt{setal07}; \object{SN~2005gl}, \citealt{galyam}), while the bipolar nebula 
surrounding the candidate-LBV \object{HD 168625} has a morphology similar 
to \object{Sk -69$^{\circ}$202}, the progenitor of SN1987A \citep{s07}. However, 
the rarity of LBVs means that the nature of these stars is still poorly understood, 
and the mechanisms behind the periodic increases in mass loss as the star leaves the 
quiescent state and the causes of the giant $\eta$-Carinae type eruptions 
are unknown. 

A recent addition to the catalogue of galactic LBVs is 
\object{W243}\footnote{RA=16~47~07.5~$\delta$=-45~52~28.5,~J2000.} 
\citep[hereafter Paper~I]{cn04} in the starburst cluster \object{Westerlund~1} 
(hereafter Wd~1; \citealt{w61, cncg05}). Early observations included 
\emph{star~G} (=\object{W243}) amongst a group of OB supergiants, with 
\cite{bks70} reporting $m_{v}$$\sim$15.1$\pm$0.2\footnote{We note that 
photometry from different studies of Wd1 tends to show systematic differences
on the order of a few tenths of a magnitude \citep{clark09} and the values 
for \object{W243} listed here may not be directly comparable.} and spectral type B0.5Ia, 
while \cite{lock74} list $m_{v}$$\sim$15.6; see also \cite{k77}.  
\cite{w87} report $m_{v}$$\sim$14.38 from photometry obtained in 1960--66, 
but used a spectrum taken in 1981 to classify \object{W243} 
as a B2Ia supergiant with a spectrum devoid of emission 
lines. However, Paper~I found that by 2002/03 \object{W243} was displaying the 
spectrum of an A2Ia supergiant, implying an apparent decrease in temperature 
of $\sim$10$^4\text{K}$ during the intervening period, accompanied by the 
development of a rich emission line spectrum showing lines of H, He~I, and 
Ca~II. Excellent agreement was found between the spectral types of six other 
B- and A-type supergiants observed by both \cite{w87} and \cite{cncg05}, and 
Paper~I concluded that \object{W243} is an LBV that was quiescent when observed 
by \cite{w87} but has since undergone a significant outburst, moving it redwards 
on the H-R diagram and (assuming constant bolometric luminosity) close to the 
Humphreys-Davidson limit. 

More recently, \object{W243} has been listed as an aperiodic variable with 
$m_{v}$$\sim$15.73 by \cite{bonanos07}, while \cite{groh06, groh07} report 
infrared spectra with Pa$\gamma$ emission that resembles the cool phase of the 
LBV \object{HR~Carinae} \citep{m02} and a K-band spectral morphology that is very 
similar to other LBVs such as \object{AG~Carinae} \citep{groh09}, \object{LBV~1806-20} \citep{eik04} and the 
\object{Pistol~Star} \citep{figer98}. However, the strong He~I emission seen 
at $6678\text{\AA}$, $7065\text{\AA}$ (Paper~I) and $10830\text{\AA}$  \citep{groh07} is 
anomalous, and \cite{groh06} note that the K-band spectrum also implies a higher 
temperature than that of a typical yellow hypergiant (YHG) and suggest that 
\object{W243} may be evolving back towards a hotter state. \cite{clark08} 
report \object{W243} as a weak (L$_{x}<10^{32}$erg s$^{-1}$) X-ray source\footnote{This 
is consistent with X-ray observations of other LBVs (e.g. \citealt{muno06}), although some 
LBVs (e.g. \object{$\eta$~Carinae}, \object{HD~5980}) have significantly 
higher X-ray luminosities (\citealt{muno06}; \citealt{clark08})}, while 
\cite{dougherty} find a current mass-loss rate for \object{W243} of 
$4-6\times10^{-6} {\rm  M}_\odot\,{\rm yr}^{-1}$ from radio observations. 
No evidence is seen of the nebulosity typically surrounding LBVs (e.g. \citealt{w03}), 
although it is likely that such a nebula would be rapidly disrupted in the 
environment of Wd~1. Finally, \cite{ritchie09} report evidence for
photospheric pulsations in \object{W243}; this is examined further in 
Section~\ref{sec:pulsations}.

This paper presents further analysis of the Paper~I dataset, along with 
analysis of five subsequent high-resolution Echelle spectra taken during 
2004 and intermediate- and high-resolution spectra from 2005, 2008 and 2009 to 
examine the physical state and evolution of \object{W243} over a seven-year 
timescale. Details of the observations and data reduction are presented in 
Section~\ref{sec:obs_data}, results are presented in Section~\ref{sec:results} 
and are discussed in Section~\ref{sec:discussion}. 


\section{Observations \& data reduction} \label{sec:obs_data}

\begin{table}
\caption{Dates of observations, instruments and configurations used. }
\label{tab:observations}
\begin{center}
\begin{tabular}{cc|ll}
Date       & MJD & Instrument & Configuration$^{a}$ \\
\hline\hline
&&\\
07/06/2002 & 52432.2 & EMMI   & REMD, 2x2, GRAT7 \\
06/06/2003 & 52796.0 & EMMI   & REMD, 2x2, GRAT7 \\
07/06/2003 & 52797.1 & EMMI   & REMD, 1x1, GRAT6 \\
21/09/2003 & 52903.0 & UVES   & RED,~580,~CD3 \\
           &         &        & RED,~860,~CD4 \\
03/04/2004 & 53098.4 & UVES   & RED,~580,~CD3 \\
           &         &        & RED,~860,~CD4 \\
           &         &        & DICHR1,~564,~CD3\\
09/06/2004 & 53165.2 & UVES   & RED,~580,~CD3 \\
           &         &        & RED,~860,~CD4 \\
           &         &        & DICHR1,~564,~CD3\\
07/07/2004 & 53193.1 & UVES   & DICHR1,~564,~CD3\\
10/07/2004 & 53196.2 & UVES   & RED,~580,~CD3 \\
           &         &        & RED,~860,~CD4 \\
           &         &        & DICHR1,~564,~CD3\\
28/09/2004 & 53276.0 & UVES   & RED,~580,~CD3 \\
           &         &        & RED,~860,~CD4 \\
           &         &        & DICHR1,~564,~CD3\\
25/03/2005 & 53454.3 & FLAMES & MEDUSA, LR6 \\
           &         &        & MEDUSA, LR8 \\
29/05/2005 & 53519.3 & FLAMES & MEDUSA, LR6 \\
           &         &        & MEDUSA, LR8 \\
13/07/2005 & 53564.1 & FLAMES & MEDUSA, LR6 \\
           &         &        & MEDUSA, LR8 \\
17/02/2006 & 53783.4 & EMMI   & RILD, 1x1, GRISM6 \\
29/06/2008 & 54646.2 & FLAMES & MEDUSA, HR21 \\
18/07/2008 & 54665.0 & FLAMES & MEDUSA, HR21 \\
24/07/2008 & 54671.1 & FLAMES & MEDUSA, HR21 \\
14/08/2008 & 54692.0 & FLAMES & MEDUSA, HR21 \\
04/09/2008 & 54713.0 & FLAMES & MEDUSA, HR21 \\
15/09/2008 & 54724.1 & FLAMES & MEDUSA, HR21 \\
19/09/2008 & 54728.1 & FLAMES & MEDUSA, HR21 \\
25/09/2008 & 54734.1 & FLAMES & MEDUSA, HR21 \\
14/05/2009 & 54965.2 & FLAMES & MEDUSA, HR21 \\
18/05/2009 & 54969.3 & FLAMES & MEDUSA, HR21 \\
\hline 
\end{tabular}
\end{center}
$^{a}$The VLT/UVES configurations list the instrument mode (RED or DICHR\#1), central wavelength (nm) 
and cross-disperser (CD\#3 or CD\#4). The NTT/EMMI configurations list the 
observing mode (REMD or RILD), binning and grating, while the VLT/FLAMES configurations 
list the GIRAFFE spectrograph mode and setup name.
\end{table}

\begin{figure*}
\begin{center}
\resizebox{\hsize}{!}{\includegraphics{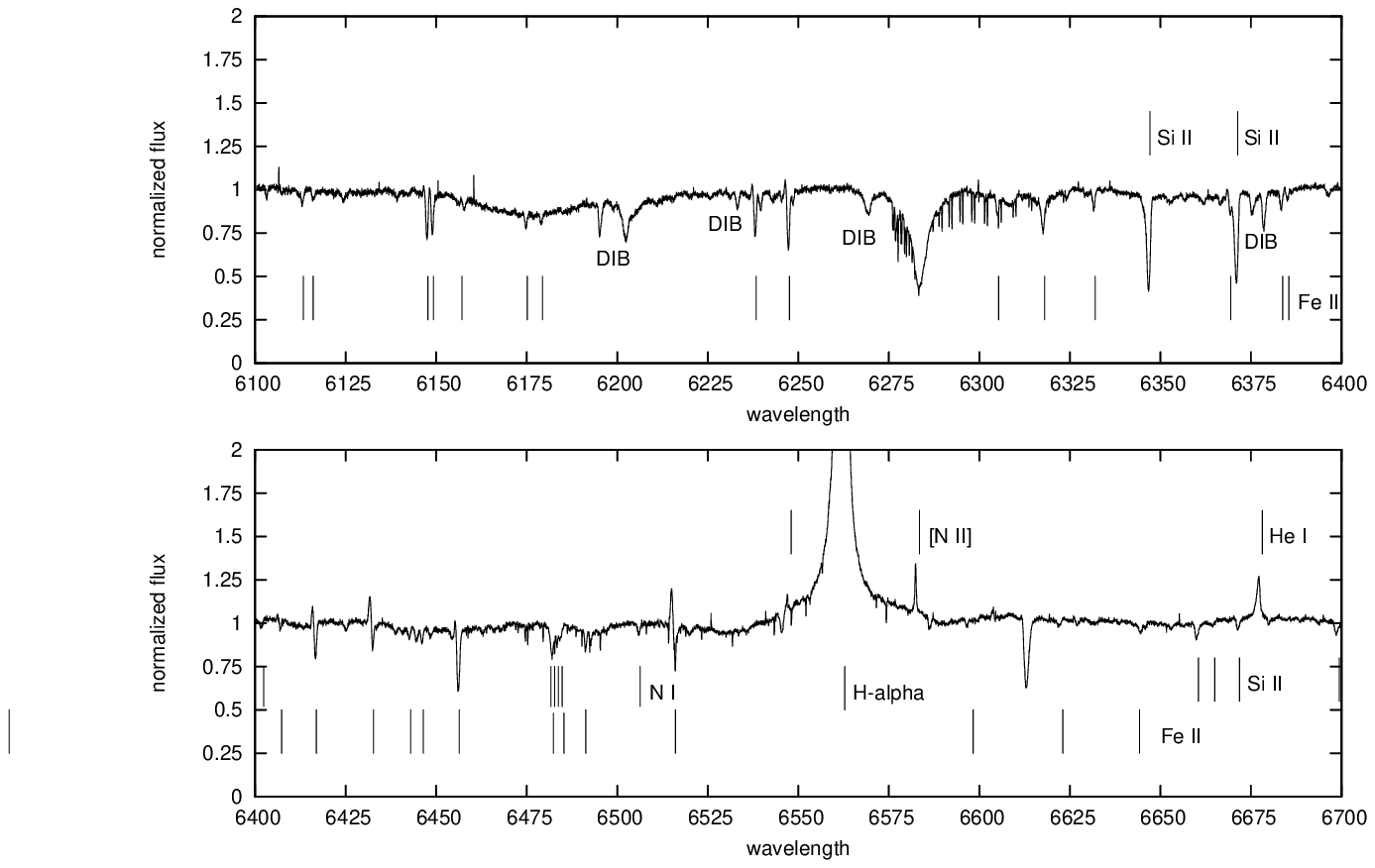}}
\caption{VLT/UVES spectrum of W243 covering $6100-6700\text{\AA}$ from 21/09/2003 (MJD=52903.0) 
with principal features discussed in the text marked. Rest wavelengths are indicated. Identified 
diffuse interstellar bands are also marked.}
\label{fig:spectrum}
\end{center}
\end{figure*}

Observations of \object{W243} were made on six nights in 2003 and 2004 using 
the Ultraviolet and Visual Echelle Spectrograph (UVES; \citealt{detal00}) 
located at the Nasmyth B focus of the 8.2m VLT UT2 \emph{Kueyen} at Cerro 
Paranal, Chile. UVES was used in RED mode with cross-dispersers CD\#3 
($4760-6840\text{\AA}$) and CD\#4 ($6600-10600\text{\AA}$), and in DICHR\#1 
mode with cross-disperser CD\#3 ($4600-6680\text{\AA}$) for the red arm. All
modes give $R \sim 40 000$. Due to the strength of the H$\alpha$ emission
from \object{W243}, two R-band spectra were taken; a 500s integration to 
capture the H$\alpha$ line, and a longer 1800s integration (2750s on 
21/09/2003, MJD=52903.0) in which the H$\alpha$ line is saturated but the 
signal-to-noise ($S$/$N$) ratio of the remaining features is improved. The $S$/$N$ ratio is 
in excess of 100 at $\lambda > 6200\text{\AA}$ for the long RED and DICHR\#1 
mode integrations, with a minimum on the UVES REDU CCD of $\sim$77 at 
$\sim$5850$\text{\AA}$, dropping below 50 at $\sim 5500\text{\AA}$ on the 
REDL CCD. The $S$/$N$ ratio averages $\sim$160 for the integrations 
centred at 860nm. Due to the high degree of reddening towards Wd~1 
($A_{v}\sim12$; \citealt{cncg05}) the $S$/$N$ ratio of data from the 
blue arm in DICHR\#1 mode (centred at 390nm) is very poor. The data were bias 
subtracted, flat fielded and wavelength calibrated using the UVES data 
reduction 
pipeline\footnote{http://www.eso.org/sci/data-processing/software/pipelines/}, 
version 3.9.0, with version 4.1.0 of the Common Pipeline Library (CPL). 

In addition to the UVES spectra, observations were made on three nights in 
2005, eight nights in 2008 and two nights in 2009 using the Fibre Large Array Multi Element 
Spectrograph (FLAMES; \citealt{pasq02}), located at the Nasmyth A focus of 
VLT UT2. FLAMES was used with the GIRAFFE spectrograph in MEDUSA mode, with 
the 2005 observations using LR6 and LR8 to cover 6438-7184$\text{\AA}$ and
8206-9400$\text{\AA}$ with $R\sim8600$ and $\sim6500$ respectively, 
and the 2008 observations using HR21 to cover the 
8484-9001$\text{\AA}$ range with $R\sim 16200$. 600s integrations were used, 
giving a $S$/$N$ ratio in excess of 100. The FLAMES data were reduced 
in a similar manner to the UVES data, using version 2.5.3 of the FLAMES-GIRAFFE 
pipeline with individual spectra extracted using the IRAF\footnote{IRAF is distributed 
by the National Optical Astronomy Observatories, which are operated by the 
Association of Universities for Research in Astronomy, Inc., under cooperative 
agreement with the National Science Foundation.} task \mbox{\emph{onedspec}}. 
Finally, we make use of ESO Multi-Mode Instrument (EMMI; \citealt{dekker86}) 
spectra covering the 8225-8900$\text{\AA}$ region from 06/2002 and 06/2003 
(see Paper~I), and a NTT/EMMI spectrum from 02/2006 covering the 
6310-7835$\text{\AA}$ region. These were obtained using the NTT at La Silla, 
Chile, and data reduction is described in Paper~I. 

A summary of all the observations is given in Table~\ref{tab:observations}. 
Analysis made use of packages within the \emph{Starlink} software suite, 
including FIGARO \citep{short97} and DIPSO \citep{how03}. Equivalent widths 
were measured using the DIPSO EW command and integrating the flux 
relative to a linear continuum based on estimated points on either side of 
the line. Radial velocities were calculated by correcting for heliocentric velocity 
and then fitting Gaussian profiles to the absorption features using IRAF
task \textit{ngaussfit}. In general this produced a satisfactory fit, but a number 
of strong lines display asymmetric wings and the measurement of 
the line centre was refined by carrying out a second fit to the core of the 
line profile, excluding the asymmetric component of the absorption line. 

\section{Description of the spectra} \label{sec:results}

The VLT/UVES and VLT/FLAMES spectra of \object{W243} reveal a complex, 
time-varying spectrum containing absorption and emission features, as well as 
absorption/emission blends with P~Cygni, inverse P~Cygni and double-peaked 
profiles: the region around the H$\alpha$ line is plotted in 
Figure~\ref{fig:spectrum}, with principal features marked. Over the entire
spectrum, strong emission is seen from the H$\alpha$ and H$\beta$ Balmer series lines, the 
Pa$\delta$~$\lambda$10049 and Pa$\epsilon$~$\lambda$9545 Paschen series lines, 
He~I~$\lambda\lambda$5015, 5876, 6678, 7065 and Ca~II~$\lambda\lambda$8498, 
8542, 8662. Weak emission is also seen from He~I~$\lambda\lambda$7281, 
8583. Forbidden line emission is seen from [N~II]~$\lambda\lambda$5754, 6584 
and [Fe~II]~$\lambda$7155, while weaker [N~II]~$\lambda6548$ and 
[Fe~II]~$\lambda\lambda$7388, 7453 lines are also seen. A possible 
[Fe~II]~$\lambda$4799 line is seen at the extreme blue end of the UVES spectra. 
Very weak [S~II]~$\lambda\lambda$6716, 6731 emission is tentatively identified. 
Many Fe~II lines are seen in both emission and absorption, with some lines 
displaying strong inverse P~Cygni profiles in the 21/09/2003 (MJD=52903.0) spectrum, 
a feature also seen in the O~I~$\lambda$8446 line and higher Paschen series lines. Strong absorption lines 
are seen from Si~II~$\lambda\lambda$6347, 6371, the O~I~$\lambda$7774 triplet 
and N~I multiplet~1 and~8 lines in the near infra-red, while many weaker 
absorption lines from singly-ionized iron group elements such as Cr~II, Sc~II 
and Ti~II are also seen: a full list is given in Table~\ref{tab:id}. Finally, the 
spectrum contains many interstellar absorption features, including diffuse 
interstellar bands, a very strong Na~I~$\lambda\lambda$5890, 5896 doublet and 
K~I~$\lambda7699$.

Due to the high degree of reddening towards Wd~1 our analysis focuses predominantly 
on features redwards of $\sim$6000$\text{\AA}$ where the $S$/$N$ ratio 
is high and spectral features are clearly defined. The short-integration UVES 
RED arm, cross-disperser \#3 spectra were used to examine the H$\alpha$ line, 
which is saturated in the longer UVES integrations. All other spectral features in 
the $4760-6650\text{\AA}$ region were examined using the long-integration 
UVES data. At longer wavelengths, UVES cross-disperser \#4 and FLAMES spectra 
were used. Unless noted in the text, rest wavelengths are taken from 
\cite{moore45} or from the NIST Atomic Spectra 
Database\footnote{http://physics.nist.gov/PhysRefData/ASD/lines\_form.html}.

\subsection{Emission lines}

\subsubsection{Balmer- and Paschen-series emission}
\label{sec:H}

\begin{table*}
\caption{Equivalent widths ($\text{\AA}$) and radial velocities (kms$^{-1}$) for the H$\alpha$, Pa$\delta$ 
and He~I~$\lambda$7065 emission lines$^{a,b}$.}
\label{tab:emissionlines}
\begin{center}
\begin{tabular}{c|ccccc|cc|cc}
     & \multicolumn{5}{c|}{H$\alpha$} & \multicolumn{2}{c|}{Pa$\delta$} & \multicolumn{2}{c}{He~I $\lambda$7065}  \\
Date       & EW    & V$_{r}$(b)& V$_{r}$ & V$_{r}$(r) & $\Delta$V$_{r}$ & EW$^{c}$ & V$_{r}$ & EW    & V$_{r}$ \\
\hline
\hline
&&&&&&&&&\\
21/09/2003 & -22.3$\pm$0.2 & -75.4     & -51.6   & -26.0      & 49.4            & -1.10$\pm$0.08    & -52.7   & -0.79$\pm$0.03 & -48.9 \\
03/04/2004 & -25.4$\pm$0.2 & -71.2     & -58.4   & -32.8      & 38.4            & -1.13$\pm$0.08    & -50.2   & -0.49$\pm$0.05 & -35.0 \\
09/06/2004 & -25.2$\pm$0.2 & -73.8     & -61.4   & -34.0      & 39.8            & -1.15$\pm$0.08    & -51.7   & -0.86$\pm$0.03 & -45.6 \\
10/07/2004 & -24.5$\pm$0.2 & -74.2     & -59.2   & -31.3      & 42.9            & -1.03$\pm$0.08    & -50.2   & -0.87$\pm$0.03 & -43.3 \\
28/09/2004 & -23.9$\pm$0.2 & -72.6     & -61.6   & -31.1      & 41.6            & -1.05$\pm$0.08    & -49.3   & -0.84$\pm$0.03 & -44.1 \\
\hline
29/05/2005 & -21.2$\pm$0.3 & --        & --      & -34.3$\pm$6& --              & --                & --      & -0.80$\pm$0.06 & -47.1 \\
\hline 
\end{tabular}
\end{center}
$^{a}$For the double-peaked H$\alpha$ line radial velocities are given 
for the bluewards and redwards peaks, V$_{r}$(b) and  V$_{r}$(r), the central absorption feature, V$_{r}$, 
and the separation of the two peaks, $\Delta$V$_{r}$.\\
$^{b}$Except where noted, errors are conservatively estimated at 
$\pm$4kms$^{-1}$ for the H$\alpha$ radial velocities and $\le \pm8$kms$^{-1}$ for the He~I~$\lambda$7065 
and Pa$\delta$ radial velocities.\\
$^{c}$ The equivalent width of the Pa$\delta$ line includes the absorption components seen in Figure~\ref{fig:HaPd} 
on either side of the emission line. If these are excluded, then the equivalent width is broadly constant at 
$\sim$1.35$\pm$0.15 $\text{\AA}$ (with increased errors due to the difficulty of estimating the continuum). \\

\end{table*}

\begin{figure}
\begin{center}
\resizebox{\hsize}{!}{\includegraphics{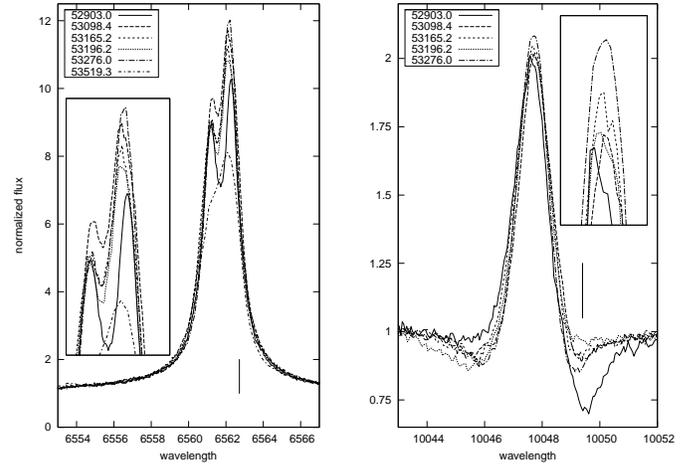}}
\caption{Evolution of the H$\alpha$ (UVES + FLAMES, left panel) and 
Pa$\delta$~$\lambda$10049 (UVES, right panel) emission lines. Insets show 
magnified views of the centre of the emission line, and rest wavelengths 
are marked in both panels.}
\label{fig:HaPd}
\end{center}
\end{figure}

Paper~I reported strengthening H$\alpha$ line emission throughout 2001-2003, a 
trend that continues into early 2004. The H$\alpha$ line, plotted in the 
left panel of Figure~\ref{fig:HaPd}, displays a double-peaked profile in all 
UVES spectra, but the central absorption feature becomes less pronounced and the 
line strengthens predominantly in the redwards peak; by mid-2004 the 
bluewards peak is at virtually the same level as the 2003 spectrum. Broad 
electron-scattering wings are visible, extending to $\pm$1000kms$^{-1}$. The 
H$\beta$ line (not shown) also strengthens in early 2004 and displays a 
similar double-peaked profile with a central absorption feature blueshifted 
50kms$^{-1}$ from its rest wavelength. Table~\ref{tab:emissionlines} lists 
the equivalent width of the H$\alpha$ line from each observation, along with 
the radial velocities of the two peaks, the central absorption feature and the 
separation between the two peaks in the profile. Lower resolution VLT/FLAMES 
spectra obtained in 2005 show a weaker but still clearly asymmetric H$\alpha$ 
line profile but cannot resolve the central absorption feature seen in the UVES spectra, 
while by 2006 a NTT/EMMI spectrum shows that the H$\alpha$ line has weakened further
to 2002/03 levels; in this case the resolution is insufficient to see if the asymmetry is 
still present. 

\begin{figure}
\begin{center}
\resizebox{\hsize}{!}{\includegraphics{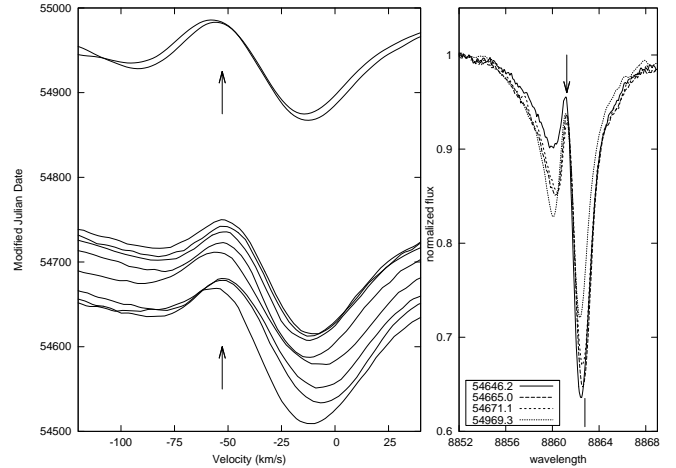}}
\caption{The Pa11~$\lambda$8863 feature over the course of the VLT/FLAMES observations. The
left panel plots the region around the line centre (with velocities relative to the Pa11 rest
wavelength) with spectra offset vertically according to the date of the observation.
The right panel overplots four spectra (three from 2008, one from 2009), with the rest wavelength 
indicated. The radial velocity of the Pa$\delta$ emission line in the 21/09/2003 (MJD=52903.2) VLT/UVES 
spectrum is marked with an arrow, showing its close correspondence with the emission feature in the 
Pa11 line and lack of long-term variability. }
\label{fig:Pavar}
\end{center}
\end{figure}

Paschen series lines are also seen in emission, with a strong 
Pa$\delta$~$\lambda$10049 line (shown in the right panel of 
Figure~\ref{fig:HaPd}) and weakening emission from the Pa8~$\lambda9546$ and 
Pa9~$\lambda9229$ lines. Unlike the Balmer-series 
lines, the Paschen-series lines do not show an obvious double-peaked profile
and the centre of the Pa$\delta$~$\lambda$10049 line is almost constant between 
observations; values are again listed in 
Table~\ref{tab:emissionlines}. The line strengthens and narrows slightly during 
2004, but the changes in equivalent width are predominantly due to the variable 
absorption on both sides of the emission line. This is redshifted relative to the 
emission line in the 2003 spectrum, but is increasingly seen on the blue side of the 
line during 2004, with a pronounced P~Cygni profile visible in the
10/07/2004 (MJD=53196.2) spectrum. The higher Paschen-series lines are seen 
predominantly in absorption, with the emission that dominates
the Pa$\delta$ line seen instead as weaker core emission that splits the blue wing
of the absorption line; this can be seen in Figure~\ref{fig:Pavar},
which plots the Pa11 line over the course of the VLT/FLAMES observations. 
The absorption component shows notable variability (discussed further in 
Section~\ref{sec:nearir}) while the emission component is largely static, 
showing little change in strength or radial velocity over the course of our
VLT/UVES and VLT/FLAMES spectra. 

\subsubsection{He~I emission}
\label{sec:He}

\begin{figure}
\begin{center}
\resizebox{\hsize}{!}{\includegraphics{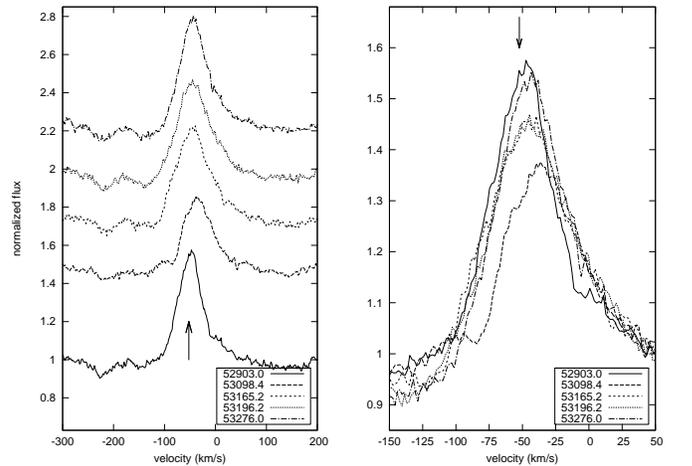}}
\caption{Evolution of the He~I~$\lambda$7065 line over the course 
of the UVES observations. Spectra are offset vertically in the left 
panel, and overplotted in the right panel. Velocities are relative to 
the rest wavelengths of the lines, and the radial velocity of the
Pa$\delta$ line in the 21/09/2003 (MJD=52903.2) UVES spectrum 
is indicated with an arrow.}
\label{fig:HeI7065}
\end{center}
\end{figure}

\begin{figure}
\begin{center}
\resizebox{\hsize}{!}{\includegraphics{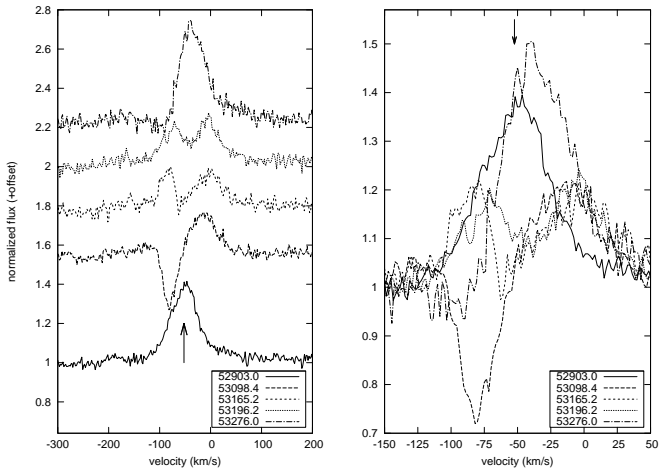}}
\caption{Evolution of the He~I~$\lambda$5876 line over the course
of the UVES observations. Spectra are shown in the same manner as
Figure~\ref{fig:HeI7065}.}
\label{fig:HeI5876}
\end{center}
\end{figure}

He~I~$\lambda\lambda$5015, 5876, 6678, 7065, 7281 and 8583 are seen in emission, 
with strong infra-red He~I~$\lambda$1.083$\mu$m and He~I~$\lambda$2.058$\mu$m 
emission lines also reported \citep{groh06,groh07}. 
No He~II features are detected. The triplet He~I~$\lambda$7065 line is plotted in 
Figure~\ref{fig:HeI7065}, with measurements of the equivalent width 
and radial velocity given in Table~\ref{tab:emissionlines}. With the 
exception of the 21/09/2003 (MJD=52903.0) spectrum, the line shows a near-constant 
profile on the red side of the emission line with emission extending 
$\sim$100kms$^{-1}$ redwards of the line centre. The bluewards emission is more
variable, decreasing substantially in the 03/04/2004 (MJD=53098.4) spectrum before
recovering to earlier levels; this decrease, which clips the peak emission, 
is responsible for the decrease in measured radial velocity listed in 
Table~\ref{tab:emissionlines}. For other epochs, the measured radial velocity
is very similar to that measured from the Pa$\delta$ line, with any differences
likely due to the blue-wing absorption. The triplet He~I$~\lambda$5876 line, shown in 
Figure~\ref{fig:HeI5876}, displays significant variability, developing a strong P~Cygni 
profile in the 03/04/2004 (MJD=53098.4) spectrum and a double-peaked profile 
separated by $\sim$80kms$^{-1}$ in the 09/06/2004 (MJD=53165.2) 
and 10/07/2004 (TJD=53196.2) spectra. The singlet He~I~$\lambda\lambda$5015, 6678
lines are also variable, although the double-peaked profile is not seen and the lines 
instead weaken to barely detectable levels before recovering in the final spectrum. 

\begin{figure}
\begin{center}
\resizebox{\hsize}{!}{\includegraphics{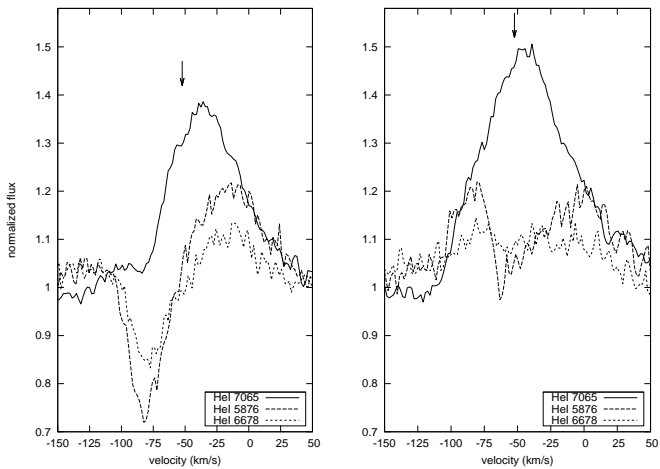}}
\caption{Comparison of the He~I~$\lambda\lambda$5876, 6678, 7065 emission lines  
on 03/04/04 (TJD=13099; left panel) and 09/06/04 (TJD=13166; right panel).
The radial velocity of the Pa$\delta$ line in the 21/09/2003 (MJD=52903.2) 
UVES spectrum is indicated with an arrow.}
\label{fig:HeIvel}
\end{center}
\end{figure}

The lowest He~I triplet term, 2s$^{3}$S, is metastable, and $n$(2s$^{3}$S) becomes 
large as the level is populated by the recombination cascade from He$^{+}$ \citep{ost89}. From here, collisional 
excitation populates the 2p$^{3}$P$^{\circ}$ upper energy level of the 
He~I~$\lambda$1.083$\mu$m line as well as the higher triplet levels (including 
the upper levels of He~I~$\lambda\lambda$5876, 7065) and the singlet levels 
responsible for the observed He~I~$\lambda\lambda$5016, 6678, 2.058$\mu$m emission 
( \citealt{bray00}; \citealt{ost89})\footnote{Absorption of $\lambda$3889 photons by the 2s$^{3}$S
level also populates the 3p$^{3}$P$^{\circ}$ level, with $\sim$10$\%$ of decays proceeding
via the 3s$^{3}$S level to 2p$^{3}$P$^{\circ}$, strengthening the He~I~$\lambda$7065 
emission line. \cite{vh72} also proposes that the 3p$^{3}$P$^{\circ}$ level 
may be populated via O~II~$\lambda$539.1 fluorescence from the He~I 1s$^{1}$S level, 
which would provide a second mechanism for strengthening the He~I~$\lambda$7065 
emission.}. The 2p$^{3}$P$^{\circ}$ level is the lower 
level of both the He~I$~\lambda$5876 and $\lambda$7065 lines, but the oscillator 
strength of the He~I~$\lambda$5876 line is almost an order of magnitude greater 
and absorption is correspondingly more pronounced; this is clearly visible in 
Figure~\ref{fig:HeIvel}, which shows the differing absorption bluewards of the two 
emission lines. Finally, we note that the weak He~I~$\lambda$8583 emission arises 
from a transition from the 10d$^{3}$D level at $\sim$24.45eV to the 
3p$^{3}$P$^{\circ}$ level \citep{vh72}. Other He~I emission lines from the high
singlet and triplet levels would be expected, e.g. the related He~I~$\lambda$8777 
line, but, given their low intensity, detection requires a combination of low-noise spectra 
and separation from other photospheric, interstellar and telluric features, and none are observed. 

\subsubsection{Fe II, Ca II and Mg I emission}
\label{sec:FeCa}

\begin{figure}
\begin{center}
\resizebox{\hsize}{!}{\includegraphics{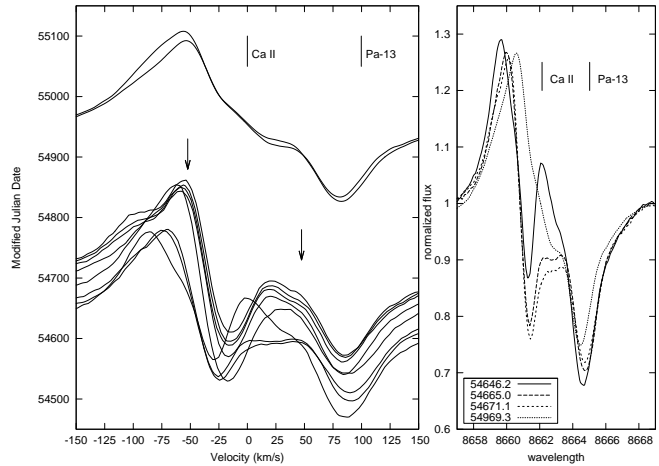}}
\caption{The evolution of the complex Ca~II~$\lambda$8662/Pa13 feaure in 
the VLT/FLAMES spectra of \object{W243}. The left panel plots the region 
around the Ca~II~$\lambda$8662 line (with velocities relative to the rest
wavelength), offset vertically according to the date of the observation. 
The rest wavelength of the Pa13 line is 
also indicated, and the radial velocity of the Pa$\delta$ line in the 
21/09/2003 (MJD=52903.2) UVES spectrum relative to the Ca~II and Pa13 rest 
wavelengths is indicated with arrows. The right panel overplots
four spectra (three from 2008, one from 2009) with rest wavelengths indicated.}
\label{fig:CaII}
\end{center}
\end{figure}

Figure~\ref{fig:CaII} shows the complex Ca~II~$\lambda$8662/Pa13 
blends over the course of ten VLT/FLAMES spectra of \object{W243}. The Ca~II emission 
comes from the multiplet-2 $4^{2}$P$^{\circ}$-$3^{2}$D transition, the upper levels 
of which are fed by the Calcium-H and -K absorption lines (which at 
$\lambda\lambda$3933, 3968 are beyond the coverage of our spectra). Unlike 
the YHG \object{IRC~+10~420} (\citealt{oud98}; \citealt{hump02}) we do not 
observe the accompanying [Ca~II]~$\lambda\lambda$7292, 7324 (multiplet 1F) 
emission that would result from a subsequent transition to the ground state; this is 
possibly masked by heavy telluric contamination, but is more probably a result of 
collisional de-excitation. In the majority of spectra the peak Ca~II emission again falls 
at the same radial velocity as the Pa$\delta$ and He~I emission, blueshifted with 
respect to a Ca~II absorption feature that is itself heavily blended with the emission 
component of the adjacent Pa13 line. However, unlike the Paschen-series lines, 
the Ca~II emission extends bluewards $\ge$100kms$^{-1}$ from the line centre and significant 
variability can be seen over short timescales in the 2008 dataset, while the central Ca~II absorption is entirely absent in the 
2009 spectra. Weak Mg~I~$\lambda$8807 and very weak 
Mg~I~$\lambda$8837 emission is also seen. These lines arise from upper energy 
levels at 5.75eV and 7.52eV, near the ionization potential of Mg~I, and are due 
to recombination of Mg$^{+}$ and the resultant cascade to the ground state. Both 
the Mg~I and Ca~II lines are likely to be formed in a neutral H~I region; the ionization 
potentials of Ca~II (11.9eV) and Mg~I (7.65eV) are too low for significant 
recombination to take place in an H~II region.

\begin{figure}
\begin{center}
\resizebox{\hsize}{!}{\includegraphics{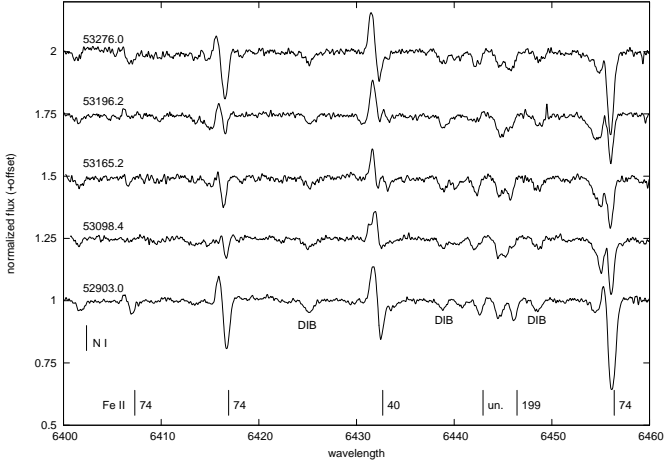}}
\caption{Evolution of the Fe~II multiplet-40 and~-74 lines over the course of
the UVES spectra. Spectra are labeled with Modified Julian Dates, and the
rest wavelengths of the principal Fe~II and N~I lines are shown.}
\label{fig:FeIIcompare}
\end{center}
\end{figure}

\begin{figure}
\begin{center}
\resizebox{\hsize}{!}{\includegraphics{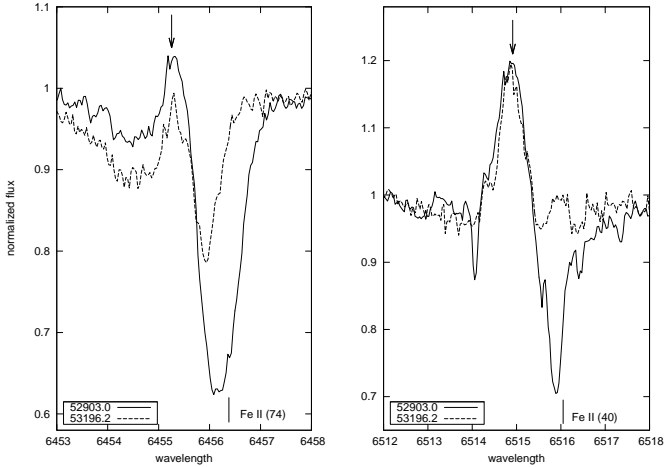}}
\caption{Comparison of the Fe~II~$\lambda$6456.4 (multiplet 74, left panel) and
Fe~II~$\lambda$6516.1 (multiplet 40, right panel) emission/absorption blends. Two 
epochs are shown, 21/09/2003 (MJD=52903.0, solid line) and
10/07/2004 (MJD=53196.2, dashed line). Rest wavelengths are marked, and
the location of the Pa$\delta$ line in the 21/09/2003 (MJD=52903.2) 
spectrum is indicated with arrows. Note the similarity between the
behaviour of the Fe~II~$\lambda$6546.4 line to the Pa11 line plotted
in Figure~\ref{fig:Pavar}.}
\label{fig:FeIIcompare2}
\end{center}
\end{figure}

Finally, a large number of Fe~II lines are also visible in Figure~\ref{fig:spectrum}. 
The strongest Fe~II features are lines from multiplets~40 (a$^{6}$S-z$^{6}$D, 
$\chi_{\text{upper}}\sim$4.8eV) and~74 (b$^{4}$D-z$^{4}$P, $\chi_{\text{upper}}\sim$5.85eV), 
which display prominent 
inverse P~Cygni profiles in the initial 21/09/2003 (MJD=52903.0) spectrum with 
emission components again displaying near-identical radial velocities to the other
emission lines in the spectrum of \object{W243}. These Fe~II lines show 
significant variations in strength and morphology; a number of these lines are plotted in 
Figures~\ref{fig:FeIIcompare}~and~\ref{fig:FeIIcompare2}, which 
show their evolution over the course of the UVES spectra. The initial inverse P~Cygni 
profiles become less pronounced and the lines of multiplet~40 are seen predominantly 
in emission, displaying notable P~Cygni profiles in the 10/07/2004 (TJD=53196.2) spectrum 
before the inverse P~Cygni profiles reform. In contrast, the multiplet~74 lines are
seen mainly in absorption, with weak emission components on the blue wing that are very
similar to the higher Paschen-series lines plotted in Figure~\ref{fig:Pavar}. 
A number of weak Fe~II lines display no obvious emission (e.g. Fe~II~$\lambda\lambda$6175, 
6332, multiplets 200 and 199 respectively). These lines all have relatively high upper 
energy levels, e.g. x$^{4}$F and 
x$^{4}$G at $\sim$8.2eV, multiplets~199 and ~200, or the unclassified multiplet containing
Fe~II~$\lambda$6644 with an upper level at $\sim$9.7eV. 
At shorter wavelengths there are emission/absorption blends due to Fe~II~$\lambda\lambda$4924, 5018, 
5169 (multiplet 42) and Fe~II~$\lambda\lambda$5235, 5276, 5316 (multiplet 49), while a large 
number of weaker Fe~II absorption lines below $6000\text{\AA}$ from multiplets 41, 46-48 
and 51-57 cannot be examined in detail due to the poor $S$/$N$ ratio.
Notably, no emission or absorption lines of either Fe~I or Fe~III 
are observed in the spectra of \object{W243}, implying that the iron must 
be almost entirely in the Fe$^{+}$ state. 

\subsubsection{[N II] and [Fe II] emission}

\begin{figure}
\begin{center}
\resizebox{\hsize}{!}{\includegraphics{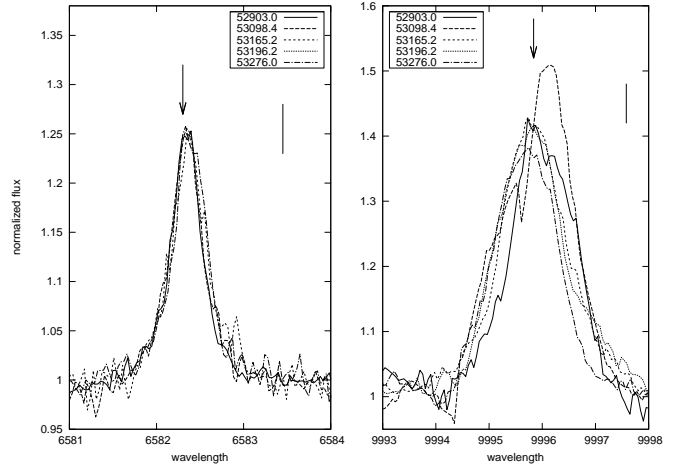}}
\caption{[N~II]~$\lambda$6583 (left panel) and the Ly$\alpha$-pumped FeII~$\lambda$9998 (right panel) emission 
lines in the UVES spectra of \object{W234}. Rest wavelengths are marked, and the location of the 
Pa$\delta$ line in the 21/09/2003 (MJD=52903.2) spectrum is indicated with an arrow.}
\label{fig:forbidden}
\end{center}
\end{figure}

The spectrum of \object{W243} displays [N~II]~$\lambda\lambda$5754, 6548, 6583 
forbidden lines (the latter plotted in Figure~\ref{fig:forbidden}), with the 
[N~II]~$\lambda6548$ line a factor of $\sim$3 weaker than [N~II]~$\lambda6583$ as expected 
from the relative radiative transition probabilities. The [N II] lines display 
virtually unchanging line profiles with a FWHM 
$\sim$25kms$^{-1}$ and near-static line centres that are blueshifted by 
$\sim$50kms$^{-1}$ from rest wavelengths. [Fe~II]~$\lambda$7155 is also seen, while 
weaker [Fe~II]~$\lambda\lambda$7388, 7453 lines are observed. At the extreme 
blue end of the 21/09/2003 (MJD=52903.0) spectrum possible [Fe~II]~$\lambda$4799 
emission is detected, but this is not observed in subsequent (lower $S$/$N$ ratio) 
spectra. In the first and last UVES spectra the [Fe~II]~$\lambda$7155 line appears very similar to the 
[N~II] emission with virtually identical FWHM and radial velocities, but the 
intermediate spectra display more complex, distorted profiles. Very weak 
[S~II]~$\lambda\lambda$6716, 6731 lines are tentatively identified.
The [Fe~II]~$\lambda$7155, 7388, 7453 lines come from 
the multiplet 14F  $a^{4}$F-a$^{2}$G transition \citep{moore45} with an upper 
energy level of 1.96eV, and the tentatively-identified [Fe~II]~$\lambda$4799 
line arises from the multiplet 4F $a^{6}$D-b$^{4}$P transition from a level 
at $\sim$2.7eV to the ground state. No higher-excitation 
lines (e.g. [Fe~III], [S~III]) are observed, implying a temperature of 
$\sim$10$^{4}$K in the line-forming region (e.g. \citealt{l88}).

\subsubsection{Ly$\alpha$ and Ly$\beta$ fluorescence emission}
\label{sec:fluorescence}

In addition to the Fe~II lines from multiplets~40 and~74, notable near infra-red emission lines 
are seen in the 21/09/2003 (MJD=52903.0) spectrum from 
Fe~II~$\lambda\lambda$8451, 8490, 8927, 9998 and O~I~$\lambda$8446. These arise from 
fluorescence by Ly$\alpha$ and Ly$\beta$ photons 
\citep{dam01}, which (like the He~I emission lines) implies the presence of a 
hot, ionizing source. The Fe~II~$\lambda$9998 line (plotted in the right panel of Figure~\ref{fig:forbidden}) 
is strong in all UVES spectra, displaying moderate variability including a notable strengthening
accompanied by a shift redwards in the 03/04/2004 (MJD=53098.4) spectrum. This 
line arises from Ly$\alpha$ pumping from the a$^{4}$G level (notably the lower level 
of the strong multiplet 46 and 49 lines seen in the spectrum of \object{W243}) to 
an upper energy level at $\sim$13.4eV, which subsequently feeds the b$^{4}$G upper level of the 
Fe~II~$\lambda$9998 line \citep{johansson84}. A weaker Fe~II~$\lambda$9956 line, produced by the same 
mechanism, is tentatively identified. \cite{sigut98} predict other fluorescence 
lines in the 8500-9200$\text{\AA}$ region, resulting from Ly$\alpha$ fluorescence 
from the a$^{4}$D level. These include Fe~II~$\lambda\lambda$8490, 8927 which are 
transitions from Ly$\alpha$-pumped $^{4}$F and $^{6}$F levels at $\sim$11.4eV:
both lines are observed in the spectrum of \object{W243}. Other predicted lines in the 
9100-9200$\text{\AA}$ range fall in a region of our spectra that is heavily 
contaminated by telluric features and are not observed. We also observe weak 
emission from Fe~II~$\lambda$8451, which shares a common upper level with 
Fe~II~$\lambda$8490 \citep{dam01}, although this is blended with a telluric 
absorption line. The majority of the fluorescence lines are only clearly detected 
in the 21/09/2003 (MJD=52903.0) spectrum, with only Fe~II~$\lambda$9998 seen strongly in all  
UVES spectra and Fe~II~$\lambda$8927 seen in the UVES and 2008 FLAMES spectra but
absent in the 2009 FLAMES spectra. The other fluorescence lines are either 
undetectable or, at best, tentatively detectable in the intermediate UVES spectra, 
and are only weakly observed in the final spectrum. 

The O~I~$\lambda$8446 line displays an inverse P~Cygni profile 
in the 21/09/2003 (MJD=52903.0) spectrum that is very similar 
to that seen in the higher Paschen-series and Fe~II multiplet-74 lines. The O~I~$\lambda$7774 triplet is 
in absorption and the O~I~$\lambda$9256 triplet, although heavily blended with a 
telluric feature, also appears to be in absorption. This suggests that the 
O~I~$\lambda$8446 emission component is a result of Ly$\beta$ fluorescence \citep{b47}; 
the pumped 3d$^{3}$D$^{\circ}$ level is linked to the upper level of O~I~$\lambda$8446 
via the permitted O~I~$\lambda$11286 line (outside the coverage of both our spectra 
and the spectrum of \citealt{groh07}), whereas the 3d$^{3}$D$^{\circ}$ level is linked 
with the upper level of the O~I~$\lambda$7774 triplet via the [O~I]~$\lambda$9205
line, which is not observed. Ionization balance calculations \citep{grandi80} show 
that Ly$\beta$ fluorescence of O~I cannot occur in either an H~I region (due to 
the lack of Ly$\beta$ photons) \emph{or} an H~II region (as the fluorescence 
mechanism requires oxygen to be predominantly neutral, and H~I and O~I have very
similar ionization potentials). Instead, the Ly$\beta$ pumping of O~I was found to 
only be effective in dense, partially-ionized transition zones between H~I and H~II 
regions \citep{kwan84}. The O~I~$\lambda$8446 line is not seen in emission \textit{or} 
absorption in the spectrum of 03/04/2004 (MJD=53098.4), recovering 
predominantly in absorption in subsequent spectra in a similar manner to the Fe~II
multiplet~74 lines discussed in Section~\ref{sec:FeCa}. 

Finally, we note that \cite{groh06} report Mg~II~$\lambda$2.14$\mu$m emission in their
spectrum of \object{W243}. This is another Ly$\beta$ fluorescence line, with pumping
from the Mg$^{+}$ ground state populating the 5p$^{2}$P$^{\circ}$ upper level of the
observed Mg~II line \citep{hs88}. 

\begin{figure*}
\begin{center}
\resizebox{\hsize}{!}{\includegraphics{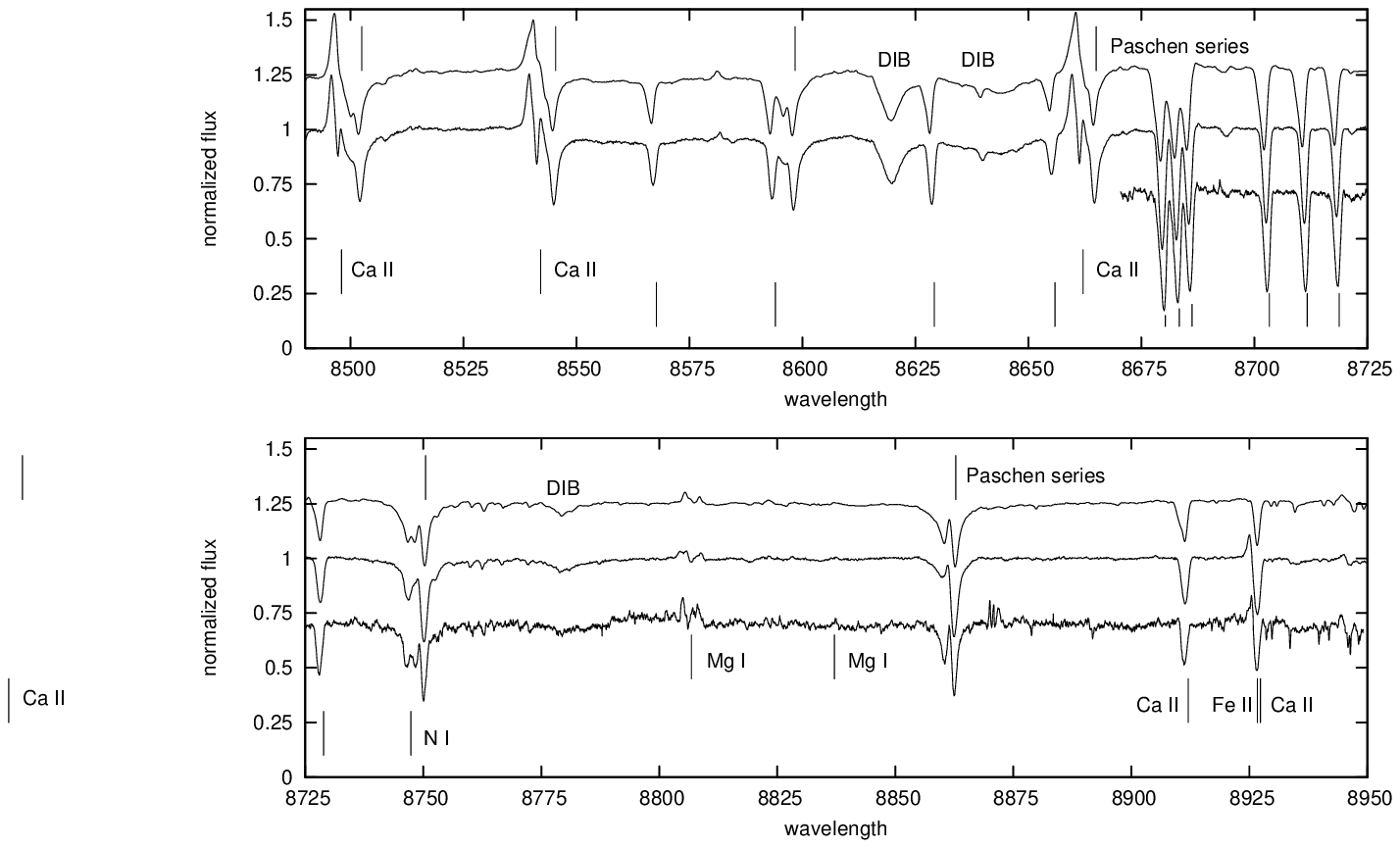}}
\caption{VLT/FLAMES (top:~18/5/2009, MJD=54969.3, middle:~29/6/2008, MJD=54646.2) and VLT/UVES 
(bottom:~28/9/2004, MJD=53276.0) spectra of \object{W243}. The rest wavelengths of principal features 
discussed in the text are indicated, and identified diffuse interstellar bands are also marked.}
\label{fig:irspectrum}
\end{center}
\end{figure*}

\subsection{Absorption lines}
\label{sec:nearir}

\subsubsection{Neutral and singly-ionized metals}

VLT/FLAMES spectra of \object{W243} from 2008 and 2009 covering 8500--8950$\text{\AA}$ 
are plotted in Figure~\ref{fig:irspectrum}, along with a VLT/UVES spectrum from 2004
covering 8670--8950$\text{\AA}$. Strong N~I absorption lines are seen 
from the 3s$^{4}$P--3p$^{4}$D$^{\circ}$ (multiplet 1) and 3s$^{2}$P--3p$^{2}$P$^{\circ}$ 
(multiplet 8) transitions, along with strong Paschen-series Pa11 \dots Pa16 absorption 
lines. Ca~II~$\lambda\lambda$8498, 8542, 8662, 8927 display emission/absorption blends,
although the Ca~II~$\lambda$8912 line is seen in absorption only; this line originates from
the same multiplet as Ca~II~$\lambda$8927, and it is likely that the emission feature in 
the latter is a blend with a Fe~II~$\lambda$8927 emission line (e$^{4}$D-5p$^{4}$D) 
that forms part of a recombination cascade from a Ly$\alpha$-pumped level at 
$\sim$10eV. Notably, this emission component is absent from the Ca~II~$\lambda$8927 line in the 2009
FLAMES spectra. 

The strong near-IR N~I absorption lines and lack of N~II features in the VLT/UVES
spectra of \object{W243} imply a spectral type no earlier than A0Ia (\citealt{mt99}, see 
also the disappearance of N~II in spectra of the LBV \object{HR~Car} as it 
cools from 20kK to 10kK; \citealt{n97} and \citealt{m02}). This is consistent with the strong 
Si~II~$\lambda\lambda$6347, 6371 absorption lines visible in Figure~\ref{fig:spectrum}, 
which are are a good indicator of temperature for cool stars. These lines are strongest at 
around 10kK (i.e. spectral type $\sim$A0), weakening as the temperature rises (\citealt{n97}; 
\citealt{davies05}). Ne~I, an additional indicator of tempratures greater than 
$\sim$10kK, is also absent. Comparison of the N~I line strength in the VLT/UVES and VLT/FLAMES 
spectra in Figure~\ref{fig:irspectrum} (along with VLT/FLAMES spectra obtained in 2005 and NTT/EMMI
spectra from 2003, neither shown) and Si~II line strengths from the VLT/UVES and 2006 NTT/EMMI 
spectra suggest that \object{W243} has shown little change in state since it finished evolving 
to the cool phase in 2002-3 (Paper~I), although we cannot rule out more significant variability 
during the periods where we do not have spectroscopic coverage. 

\subsubsection{Pulsations}
\label{sec:pulsations}

\begin{figure}
\begin{center}
\resizebox{\hsize}{!}{\includegraphics{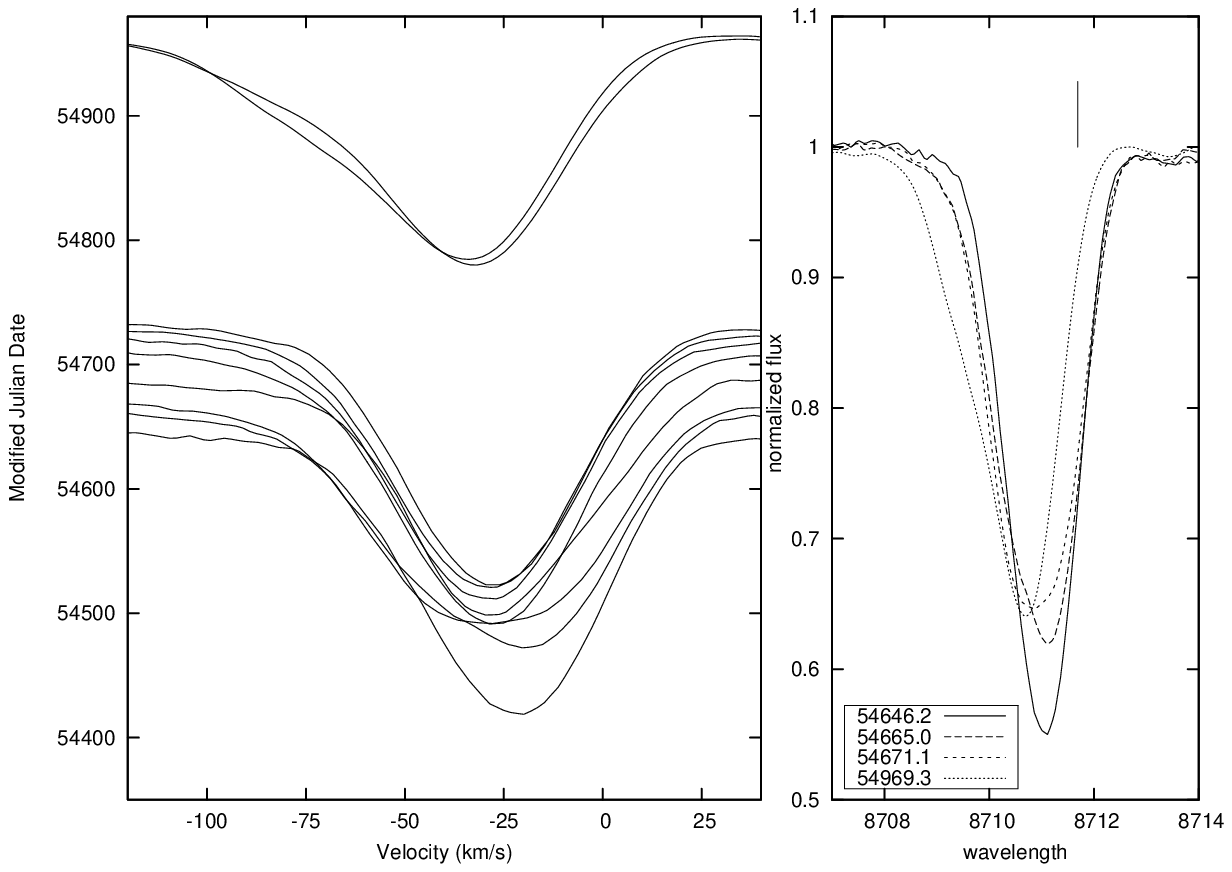}}
\caption{Evolution of the N~I~$\lambda$8711 absorption lines during the VLT/FLAMES 
observations. As in Figures~\ref{fig:Pavar} and~\ref{fig:CaII}, the left panel 
plots the region around the line centre, with velocities relative to the rest 
wavelength, for all VLT/FLAMES spectra, offset vertically according to the date of
observation. The variability in line centre and profile can be clearly seen, with
a broad blue wing extending to $\ge$100kms$^{-1}$ in the 2009 observations. The
right panel overplots four spectra (three from 2008, one from 2009) with 
the rest wavelength indicated.}
\label{fig:NIvar}
\end{center}
\end{figure}

\begin{figure}
\begin{center}
\resizebox{\hsize}{!}{\includegraphics{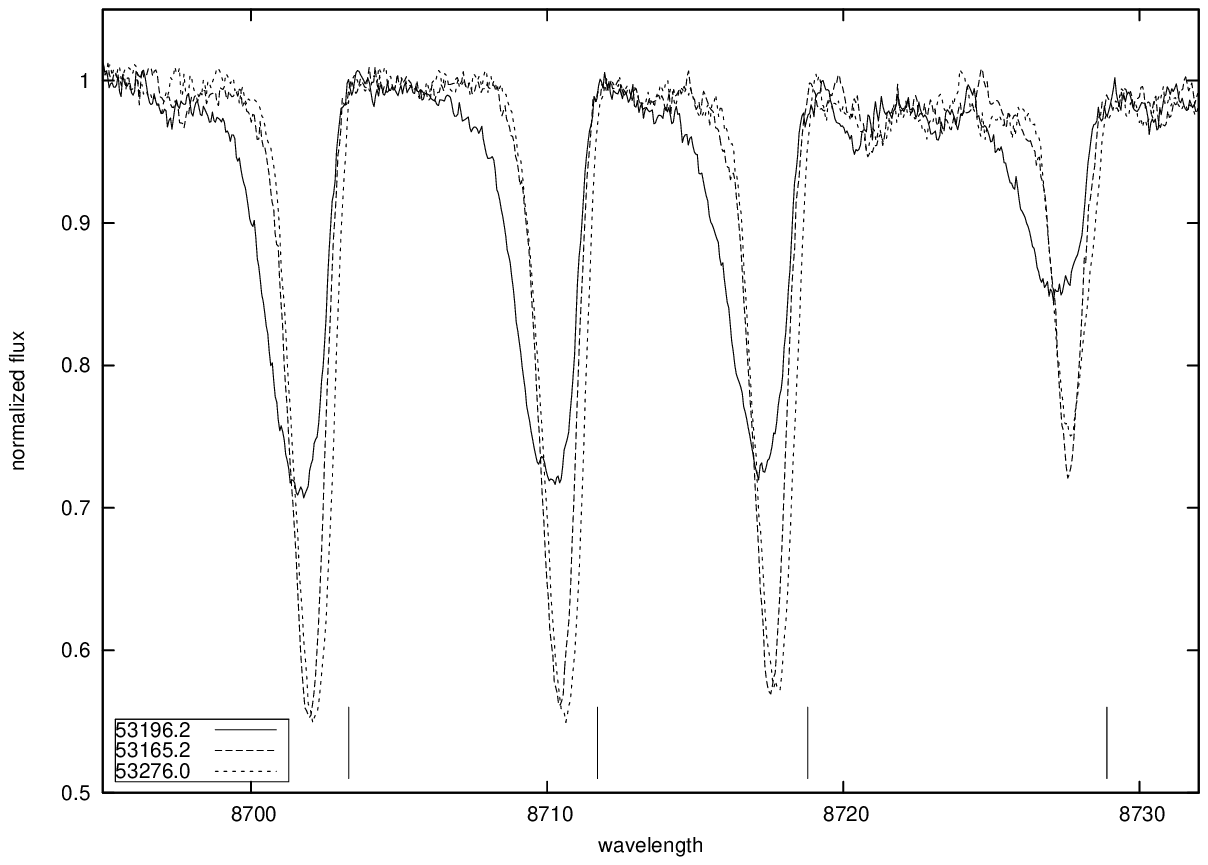}}
\caption{UVES observations of the N~I absorption line complex around 
$\sim$8700$\text{\AA}$. on 10/07/2004 (MJD=53196.2, solid line) with 
preceeding (09/06/2004, MJD=53165.2) and subseqent (28/09/2004, MJD=53276.0)
spectra plotted with dashed lines for comparison. Rest wavelengths are marked.}
\label{fig:NIoutburst}
\end{center}
\end{figure}

Figure~\ref{fig:NIvar} shows the N~I~$\lambda$8711 absorption line from the 
2008 and 2009 VLT/FLAMES datasets, with spectra offset vertically according to their acquisition
date to highlight changes in the line profiles. The line displays significant variability 
in strength,  profile and line centre, and similar variations are seen in the Pa11 line, plotted in 
Figure~\ref{fig:Pavar}, the Pa13 line visible in Figure~\ref{fig:CaII} and the 
strong Si~II~$\lambda\lambda$6347,6371 doublet and O~I~$\lambda$7774 triplet lines (not shown).  
Over the course of sixteen epochs of VLT/UVES 
and VLT/FLAMES data the measured radial velocities from the strong neutral/singly-ionized
metal lines lie within a range of -43kms$^{-1}$ -- -18kms$^{-1}$ (errors $\le\pm3$kms$^{-1}$)
with individual lines consistent to within a few kms$^{-1}$ in each spectrum; only the 
low-excitation Fe~II multiplets displaying P~Cygni and inverse P~Cygni profiles 
in Figure~\ref{fig:FeIIcompare} are discrepant, displaying red-shifted absorption 
components relative to the other metal lines, most probably as a result of significant 
infilling from the blue-wing emission. 

As noted in \cite{ritchie09}, the short-term radial velocity variations 
are hard to reconcile with orbital motion, as the implied orbital period is at odds
with the narrow range of measured radial velocities unless we are viewing the system at
\textit{sin i}$\sim$0. In addition, the clear changes in line profile and strength
apparent in Figure~\ref{fig:NIvar} are hard to understand in this scenario,
and we instead interpret the radial velocity variations as resulting from bulk motions 
in a pulsating, dynamically-unstable photosphere. Such pulsations are well-documented 
in YHGs, for example \object{$\rho$~Cassiopaeia} \citep{lobel98}, and very similar changes 
in N~I and Fe~I absorption lines are seen in VLT/FLAMES spectra of the pulsating YHG 
\object{W265} \citep{ritchie09}. The limited spectral coverage and lack of 
contemporaneous photometry makes determination of a pulsation period for \object{W243}
impossible; the behaviour of the N~I lines in the VLT/FLAMES spectra
is consistent with a long pulsational period  similar to that observed in YHGs 
\citep{deJager98}, but the photospheric lines also display rapid changes in both radial
velocity and profile over a matter of days (e.g. the change between 18/07/2008, MJD=54665.0
and 24/07/2008, MJD=54671.1 visible in Figures~\ref{fig:Pavar} and~\ref{fig:NIvar}) that 
could imply that we are observing the effects of sparse sampling of a more rapid
pulsational period. 

A further similarity with the YHGs \object{$\rho$~Cas} and \object{W265} comes from the
weakening and broadening of the metal absorption lines through the pulsation period, and in
particular the development of the broad blue wing seen in the 
near-IR N~I lines on 10/07/2004 (MJD=53196.2, see Figure~\ref{fig:NIoutburst}); the O~I and Si~II absorption lines 
display a similar effect. The observed line broadening 
appears periodic, and is also seen in the N~I lines in the 18/05/2009 (MJD=54969.3) VLT/FLAMES 
spectrum plotted in Figure~\ref{fig:irspectrum}, most clearly in the N~I and Paschen-series absorption 
lines and the significant reduction in core emission in the N~I triplet at $\sim$8680$\text{\AA}$.
In \object{$\rho$~Cas} this behaviour is correlated with maximum effective temperature \citep{lobel98}, 
and results from absorption in the wind due to a periodic phase of enhanced mass-loss\footnote{Note that
this is not the \textit{strongly enhanced} mass loss phase encountered by YHGs on timescales of a few 
decades, accompanied by a significant movement redwards on the HRD (\citealt{deJager98};\citealt{lobel03})}. 
\object{W243} therefore appears to be in a relatively quiescent pulsational state similar to warm-phase 
YHGs, with possible periods of increased mass-loss but no significant changes in state apparent
during our monitoring.

\subsubsection{Modeling}

\begin{table}
\caption{Model parameters for \object{W243}. }
\label{tab:model}
\begin{center}
\begin{tabular}{ll}
Parameter        & Value$^{a}$ \\
\hline\hline
&\\
$L_{*}$          & $7.3\times10^{5}~(d/4.5\text{kpc})^{2}$L$_{\odot}$  \\  
$R_{*}$          & $376.9~(d/4.5\text{kpc})$R$_{\odot}$ \\
$R_{2/3}$        & $450.0~(d/4.5\text{kpc})$R$_{\odot}$ \\
$T_{\text{eff}}$ & $\sim$8500K$^{b}$\\
$\dot{M}$        & $6.1\times10^{-7}~(d/4.5\text{kpc})^{1.5}$~M$_{\odot}$yr$^{-1}$ \\
$V_{\infty}$     & 165kms$^{-1}$\\
$v~\text{sin}i$  & 10kms$^{-1}$\\
log~$g$          & $\sim$0.65$^{c}$ \\
H/He             & 5\\
N/N$_{\odot}$    & 12.1\\
O/O$_{\odot}$    & 0.11\\
Ca/Ca$_{\odot}$  & 2.49\\
Mg/Mg$_{\odot}$  & 1.34\\
Si/Si$_{\odot}$  & 1.51\\
Fe/Fe$_{\odot}$  & 1.22\\
\hline 
\end{tabular}
\end{center}
$^{a}$Values for $V_{\infty}$, $v~\text{sin}i$ and H/He are assumed, all others are derived. Values are for 
a distance of 4.5kpc \citep{cncg05,crowther06}, with scaling listed where appropriate.\\
$^{b}$Limits are 8300$\le T_{\text{eff}}\le$9300K, favouring a value at the lower end of this range.\\ 
$^{c}$Best fit to the higher Paschen-series absorption lines.\\
\end{table}

\begin{figure}
\begin{center}
\resizebox{\hsize}{!}{\includegraphics{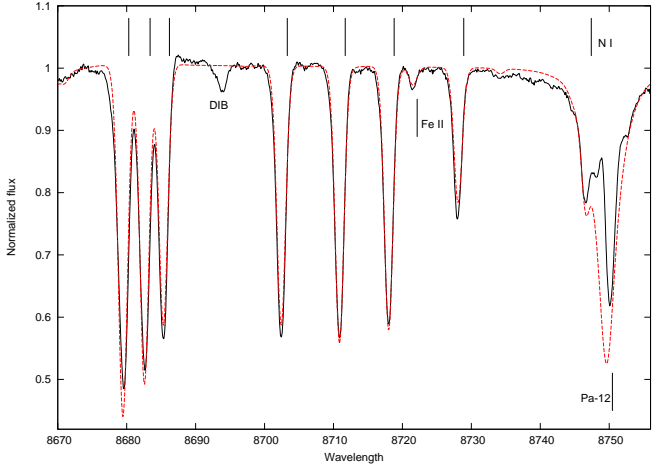}}
\caption{Comparison of the model fits to the strong N~I multiplet~1 absorption lines and
adjacent Pa11~$\lambda$8750 line. The VLT/FLAMES spectrum is from 04/09/2008 (MJD=54713.0) 
and rest wavelengths are marked.}
\label{fig:model1}
\end{center}
\end{figure}

\begin{figure}
\begin{center}
\resizebox{\hsize}{!}{\includegraphics{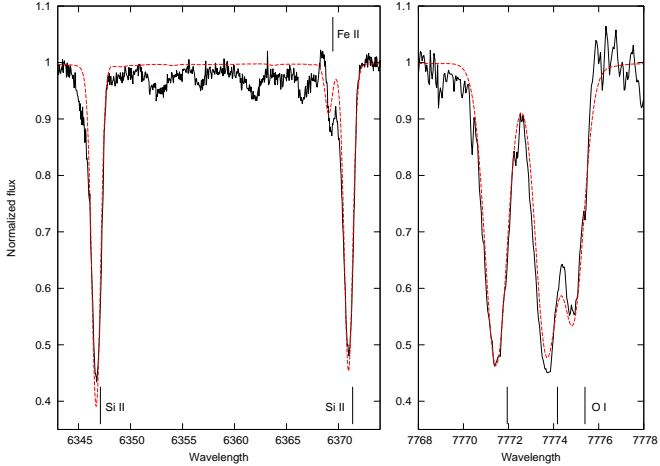}}
\caption{Model fits to the Si~II~$\lambda\lambda$6347, 6371 doublet (left panel) and 
O~I~$\lambda$7774 triplet (right panel). The VLT/UVES spectrum is from 21/09/2003 (MJD=52903.0) and 
rest wavelengths are marked.}
\label{fig:model2}
\end{center}
\end{figure}

\begin{figure}
\begin{center}
\resizebox{\hsize}{!}{\includegraphics{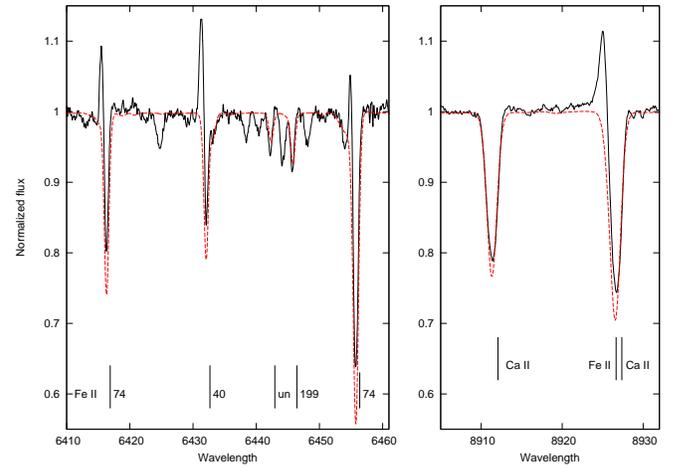}}
\caption{Example of model fits to emission/absorption blends. The left panel shows three
strong Fe~II multiplet~40 and~74 lines, while the right panel shows the 
Ca~II~$\lambda\lambda$8912, 8927 doublet and superimposed Fe~II~$\lambda$8927 emission 
line. In both cases the absorption component is well fit by the model, but the cool
star cannot reproduce the emission component. The VLT/UVES spectrum is from 21/09/2003 (MJD=52903.0)
and the VLT/FLAMES spectrum is from 04/09/2008 (MJD=54713.0).
Rest wavelengths are marked, and the Fe~II multiplets are also indicated in the left panel.}
\label{fig:model3}
\end{center}
\end{figure}

Modeling of \object{W243} was carried out using the non-LTE line-blanketed radiative transfer code 
CMFGEN (\citealt{hillier98}; \citealt{hillier99}), which computes line and continuum formation 
under the assumptions of spherical symmetry and a steady-state outflow. Examples of the use of
CMFGEN to model LBVs and detailed discussion of the workings of the code can be found in \cite{najarro09} 
and \cite{groh09} and extensive references therein. For \object{W243}, the H/He abundance was fixed at 5, 
while C was assumed to be depleted due to the lack of observed lines. Ti, Cr and Ni have been assumed 
to have solar abundance, whereas abundance determinations for Fe, Ca, Mg, Si, N and O are firmly 
derived. We find that the absorption-line spectrum of \object{W243} is well fit by a cool hypergiant model 
with a weak, unclumped wind\footnote{For the velocity law we assume $V_{\infty}$=165kms$^{-1}$, appropriate for a quiescent YHG, 
and $\beta$=1} and low mass-loss rate. 
Model parameters are given in Table~\ref{tab:model}; a distance of 4.5kpc \citep{cncg05, crowther06} 
is assumed, with distance-scaling relationships given as appropriate \citep{hillier98b}. The value of log~$g$ 
is uncertain, with the listed value of $\sim$0.65 providing the best fit to the relatively uncontaminated higher
Paschen-series lines bluewards of 8500$\text{\AA}$. Figure~\ref{fig:model1} shows the 
model fit to the strong N~I multiplet~1 lines and the adjacent Pa12~$\lambda$8750 line (blended 
with N~I~$\lambda$8747), while Figure~\ref{fig:model2} shows the fit to the strong 
Si~II~$\lambda$6347, 6371 doublet and O~I~$\lambda$7774 triplet lines. All of the strong metal absorption 
lines are well fit by the model, although the Pa12 line shows significant infilling from the emission 
component discussed in Section~\ref{sec:H}. Figure~\ref{fig:model3} shows three strong Fe~II multiplet~40 
and~74 lines (two weaker lines from multiplet~199 and a multiplet unclassified by \cite{moore45} also visible, along
with several weak DIBs), along with the Ca~II~$\lambda\lambda$8912, 8927 doublet and overlapping 
Fe~II~$\lambda$8927 emission line. In both cases the model reproduces the absorption component 
well, but does not reproduce the emission components. Similarly, the model predicts Paschen- 
and Balmer-series lines also in absorption and very weak He~I absorption features, rather than
the observed emission. Along with the Ly$\alpha$-pumped Fe~II and O~I emission lines, this
implies that the emission-line spectrum does not originate with the cool star but 
instead has a secondary source; this is discussed further in Section~\ref{sec:discussion}. 

A strong lower limit on temperature come from the Mg~I~$\lambda$8807 line,
while upper limits come from the He~I and N~I lines (the strength of the latter also used as a 
temperature diagnostic in Paper~I). Precise temperature determination is complicated by
the presence of emission components in both He~I and Mg~I, and we find 8300$\le T_{\text{eff}}\le$9300K,
(i.e. spectral type A2--4Ia$^{+}$), favouring a value of $\sim$8500K (A3Ia$^{+}$). At lower 
temperatures the Mg~I absorption component would become too strong, even considering infilling 
from the superimposed emission visible in Figure~\ref{fig:irspectrum}, while at higher temperatures 
the unblended N~I absorption lines become too weak. N is highly overabundant (12.1$\times$~solar), with 
O depleted (0.11$\times$~solar) and C highly depleted. This is clearly indicative of CNO-processed material
being present at the surface of \object{W243}, with abundances similar to those determined
for \object{AG~Car} \citep{groh09}. Ca is also notably overabundant, with Mg, Si and Fe all moderately 
enhanced relative to solar values.  

\section{Discussion and Conclusions}\label{sec:discussion}

\subsection{The spectrum of W243}

We summarize the spectrum of \object{W243} as follows:
\begin{itemize}
\item \textbf{H, He~I, O~I, Mg~I, Ca~II, Fe~II, [N~II] and [Fe~II]} are seen in emission. Strong H$\alpha$, H$\beta$ and 
Pa$\delta$ emission is seen, with the H$\alpha$ and H$\beta$ lines displaying double-peaked profiles, 
while the higher Paschen-series lines show core emission on the blue absorption wing. He~I emission is 
highly variable, while O~I $\lambda$8446 and many Fe~II lines also show significant variability. 
[N~II] and a few [Fe~II] lines are observed. Apart from a few highly-distorted lines, all permitted and forbidden 
emission lines display a blueshift of $\sim$50kms$^{-1}$ with negligible epoch-to-epoch variability. 

\item \textbf{Many neutral or singly-ionized metals are seen in absorption}. Strong absorption lines of 
Si~II, N~I and O~I are seen, along with many Fe~II lines; Fe~I and Fe~III features are not observed. 
Many weaker lines from singly-ionized iron-group elements are also seen. Modeling of the absorption line
spectrum shows 8300$\le$T$_{\text{eff}}\le$9300K (A2--A4Ia$^{+}$, with $T_{\text{eff}}\sim$8500K preferred). 
This is strongly inconsistent with the observed He~I emission and Ly$\alpha$ and Ly$\beta$ fluorescence 
lines of O~I and Fe~II, implying the presence of an unseen source of ionizing photons. 

\item \textbf{Photospheric pulsations are apparent in the metal absorption lines},
with periodic broadening and the development of excess blue-wing absorption suggestive of periods of increased
mass loss. However, no evidence is seen for eruptive mass loss, and \object{W243} appears to be in a
quiescent cool-phase state.
\end{itemize}

\subsection{The current state of W243}

With $L_{*}$=7.3$\times$10$^{5}L_{\odot}$ (assuming $d$=4.5kpc; \citealt{cncg05}) and $T_{\text{eff}}$$\sim$8500K, \object{W243} lies near the 
post-Main Sequence track of a $\sim$40$M_{\odot}$ star (e.g. \citealt{schaller92}), falling as expected 
between the $\sim$40-55$M_{\odot}$ progenitors of the Wd1 Wolf-Rayet population \citep{crowther06} and 
the $\le$30$M_{\odot}$ $\sim$O7-8V stars at the main sequence turn-off \citep{negueruela09}. 
\object{W243} appeared on the blue side of the HRD as a luminous, early-B supergiant when observed by \cite{bks70}, 
\cite{lock74} and \cite{w87} before undergoing an (unobserved) event that led to \object{W243} moving redwards 
to the `yellow void' and taking on its current appearance as an early-A hypergiant.
The final stages of this evolution were seen in the early observations described in Paper~I, with the star remaining 
generally quiescent in its cool-phase state since 2003. Our modeling shows 
Nitrogen to be highly overabundant, with the depletion of Carbon and Oxygen suggesting we are observing 
significant quantities of CNO-processed material. Other $\alpha$-process elements (Ca, Mg, Si) are
also enhanced. The implication is therefore that \object{W243} is either in an
advanced pre-RSG LBV phase, having already ejected its less processed, Hydrogen-rich outer layers, or has evolved 
through the RSG phase and returned to the blue side of the HRD, with the current CNO-rich material dredged up
during the RSG phase. \object{W243} does not display the extended circumstellar ejecta expected for \textit{either} 
evolutionary stage, but, as commented on in Paper~I, this is likely to be rapidly dispersed in the extreme environment of Wd1.
While the N/O ratio in \object{W243} is similar to \object{AG~Car} \citep{groh09}, the latter object is both more massive 
and (considerably) more luminous than \object{W243} and the same evolutionary phase cannot be assumed, while
post-RSG objects such as \object{IRC~+10~420} are also overabundant in Nitrogen \citep{oudmaijer}\footnote{Sodium 
abundances are used to infer a post-RSG phase for \object{$\rho$~Cas} 
(e.g. \citealt{deJager97}), but the only features apparent in our spectra result from interstellar absorption.}. 
However, future abundance studies of the unique coeval population of early-B supergiants, YHGs and BIa$^{+}$/WNVL Wolf-Rayet 
precursors found in Wd1 offer the opportunity to accurately constrain both the evolutionary state of \object{W243} itself
and the general passage of massive stars as they leave the main sequence and evolve through the `zoo' of transitional objects  
towards the Wolf-Rayet phase. 

\subsection{Comparison with other stars}

\subsubsection{Cool-phase LBVs}

The visual and near-IR spectrum of \object{W243} shows many similarities with other cool-phase LBVs.
The Si~II doublet is prominent in cool supergiants \citep{davies05},
and is therefore reported in many objects, while the strong near-IR N~I lines are also seen at $\sim$B8Ia and later, 
e.g. in the spectra of \object{S~Dor} \citep{munari09} and \object{HR~Car}\footnote{The N~I lines are notably stronger in \object{W243} than in 
\object{HR~Car}, implying a later spectral type \citep{mt99}.} \citep{m02}. The R-band spectrum of 
\object{W243} shows strong similarities with the A2Ia$^{+}$ LBV \object{HD~160529} \citep{stahl03, chentsov03}, although
subtle differences are present, with the C~II~$\lambda\lambda$6578, 6583 doublet notable in the spectrum of \object{HD~160529}
and Ne~I also reported; neither can be robustly identified in the case of \object{W243}. The rich Fe~II spectrum  
is also a typical feature of cool-phase LBVs, with \object{HR~Car} a canonical
example, displaying prominent Fe~II emission/absorption blends in the cool states of 1991 and 1998--2002 that fade 
as the LBV moves bluewards on the HRD. The transition from P~Cygni to inverse P~Cygni profiles in Fe~II seen in
Figure~\ref{fig:FeIIcompare} is also reported in the cases of \object{S~Dor} \citep{wolf90, wolf92}, \object{R127}
\citep{wolf92, walborn08} and \object{HR~Car} (\citealt{n97}; \citealt{m02}; Crowther, priv. comm. 2009); we return to this issue in Section~\ref{sec:location}.

However, notable differences also exist between \object{W243} and other early-A LBVs. The H$\alpha$ 
and H$\beta$ lines seen in the spectrum of \object{W243} are unusual, appearing strongly in emission but lacking the
P~Cygni absorption component generally seen in LBV spectra. The strong He~I and 
Ly$\alpha$-pumped Fe~II and O~I emission in \object{W243} are also clearly discrepant with the cool-phase
emission and absorption lines and are more in keeping with early-B/Ofpe hot-phase LBVs (e.g. \object{AG~Car}; \citealt{groh09}),
although other characteristic hot-phase features such as emission lines of Fe~III, N~II, Si~II and Si~III are absent. 
The He~I~$\lambda\lambda$5876, 6678 lines are seen in absorption in the cool-phase spectra of \object{HR~Car} \citep{m02}, 
and the four B6-A2 (candidate) LBVs described by \cite{chentsov03}, while He~I lines
appear in emission in \object{R127} only as the Fe~II spectrum fades during the transition to the hot, quiescent state 
\citep{walborn08}. \object{HD~160529} provides a striking example of these differences, with the otherwise-similar 
A2Ia$^{+}$ spectrum showing He~I in absorption and a strong P~Cygni profile in the H$\alpha$ and H$\beta$ lines;
our model also suggests that He~I should be visible weakly in absorption, rather than the observed 
strong emission.
Infra-red spectra of the mid-B or later (candidate) LBVs listed by \cite{groh07} again show \object{W243}
to be atypical, with a strong He~I~$\lambda$1.083$\mu$m emission line (of slightly greater strength than the adjacent
Pa$\gamma$ line) contrasting with the remainder of the sample that show the line to be in absorption in two cases 
(\object{HD~168625}, \object{HD~168607}), almost absent in \object{HR~Car} and showing a weak emission component with strong 
P~Cygni absorption in \object{HD~160529}. 

\subsubsection{Yellow Hypergiants}

Our modeling shows that the absorption line spectrum of \object{W243} is well described by a cool hypergiant close 
to the Eddington limit, and we observe spectral variability that is similar to the well-studied YHGs \object{$\rho$~Cas} 
\citep{lobel98} and \object{HR~8752} \citep{deJager97}. Within Wd1, \object{W265} also displays pulsational variability in 
the near-IR N~I lines, while at radio wavelengths \object{W243} is very similar to the F2Ia$^{+}$ YHG \object{W4} \citep{dougherty}. 
The lack of Fe~I features indicates that \object{W243} is hotter than these objects, which show strong neutral metal absorption and 
a complete absence of He~I, with the latter also not observed in the A-type YHGs \object{W12a} (A5Ia$^{+}$) and \object{W16a} 
(A2Ia$^{+}$). The peculiar mid-A YHG \object{IRC~+10~420} (\citealt{oud98}; \citealt{hump02}) 
shows the strongest spectral similarities to \object{W243}, with Fe~II multiplet~40 and ~74 lines in emission and
a strong, double-peaked H$\alpha$ emission line with broad electron-scattering wings; both 
objects also display double-peaked Ca~II emission. However, the close agreement between the H$\alpha$ and Ca~II 
features in velocity space (e.g. Figure 5, \citealt{hump02}) is not seen in \object{W243}\footnote{This may be 
a result of the blending with the adjacent Paschen-series absorption/emission features which are not seen strongly 
in \object{IRC~+10~420}}, suggesting that the origin may not be the same. In addition, the 2009 VLT/FLAMES spectra 
show that the Ca~II infra-red triplet no longer displays a double-peaked profile, with only a weak `step' in the
red flank of the emission line hinting at the previous location of the absorption feature. Lack of 
contemporaneous R-band spectra means that we cannot tell if the double-peaked H$\alpha$ profile is also now absent, 
although we note the clear weakening in the redwards peak in the VLT/UVES spectra and its apparent absence in the 
lower-resolution VLT/FLAMES LR6 spectra from 2005 (see Figure~\ref{fig:HaPd}). Therefore, although the double-peaked
line profiles seen in \object{W243} may correspond to the disk (e.g. \citealt{jones93}) or bipolar outflow 
\citep{oud94, davies07} models proposed for \object{IRC~+10~420}\footnote{\cite{hump02} also propose a complex 
\textit{rain} model for \object{IRC~+10~420} in which both inflow and outflow are present in an opaque wind. However,  
this requires a mass-loss rate $\sim$10$^{3}$ times greater than currently appears to be the case for \object{W243}.}, 
the central absorption may also result from radiative transfer effects in the line-forming region; additional modeling 
would be required to confirm this. 

\subsection{Origin of the emission line spectrum}

While the spectrum of \object{W243} shows similarities with both cool, A-type LBVs and A-~and~F-type YHGs, 
the juxtaposition of a cool hypergiant absorption line spectrum with hot-phase LBV He~I and Lyman-pumped
Fe~II and O~I emission lines is an obvious discrepancy. However, our model is 
unable to reproduce \textit{any} of the observed emission features while simultaneously preserving the N~I and 
Si~II absorption lines, and we interpret this as an indication that \object{W243} is a binary, consisting of the 
cool-phase LBV and an undetected hot OB (or possibly Wolf-Rayet) companion that is the source of the ionizing flux responsible 
for the observed radio emission and discrepant emission lines. The binary fraction amongst massive stars in Wd~1 is known to be large 
\citep{crowther06, clark08, ritchie09}, and although \object{W243} lacks the strong 
non-thermal radio and X-ray emission that provide evidence for colliding winds in a binary system, weak X-ray emission 
is observed \citep{clark08}. Despite the implied presence of a hot companion to the LBV, radial velocity measurements are not obviously 
consistent with binarity unless we are viewing \object{W243} at a highly favourable angle or the companion is in a long-period, 
possibly highly-eccentric orbit around the LBV primary as has been suggested for \object{$\eta$~Carinae} \citep{nielsen07}. 
An alternative configuration may be similar to that observed in the YHG \object{HR~8752}, where the flux from a B1V companion 
in a wide orbit is responsible for the formation of the observed [N~II] emission and a compact radio nebula, neither
of which could be formed by the cool hypergiant itself \citep{stickland78}. A similar scenario may also apply to the YHG 
\object{W265}, which also displays [N~II] and radio emission \citep{ritchie09}. In the case of \object{W243} a hotter 
companion is probably required to provide the ionizing flux necessary to form the observed He~I and Ly$\alpha$-pumped 
emission, both absent in \object{HR~8752} and \object{W265}, although we note the absence of higher-excitation emission 
lines such as N~II, [S~III], [Fe~III] and [Ar~III] that are observed in the sgB[e] binary 
\object{W9} \citep{cncg05, clark08}, which also displays far stronger Ly$\alpha$-pumped lines than those observed in \object{W243}. 

The lack of observed Fe~I or Fe~III features in the spectrum of \object{W243} is unusual and implies that almost all of the iron 
is in the Fe$^{+}$ state in the line formation region. This is reminiscent of the B and D~Weigelt blobs of \object{$\eta$~Carinae}
\citep{verner02}, which display spectra dominated by Fe~II and [Fe~II] emission originating from upper levels 
around 6eV (permitted lines) and 2eV (forbidden lines). This Fe~II spectrum arises 
from collisional excitation at $10^{6}<n_{e}<10^{7}$cm$^{-3}$ populating the lower Fe~II levels, with strong continuum 
pumping routes to $\sim$6eV levels arising from the a$^{4}$D and a$^{2}$G levels (the a$^{2}$G level is the 
upper level of our observed [Fe~II] multiplet 14F emission) and Ly$\alpha$ pumping from the a$^{4}$D and a$^{4}$G 
levels; this also appears a plausible model for the Fe~II emission region of \object{W243}\footnote{The similarity between the
Fe~II-rich spectra of \object{W243} and other cool-phase LBVs, e.g. \object{HR~Car} or \object{S~Dor}, also suggests similar 
conditions in the Fe~II line formation regions, although we do not suggest that a hot companion is required in all of these 
objects; instead, the similarity probably reflects the complex balance of collisional, continuum and Ly$\alpha$-pumping processes 
in these objects \citep{verner02}.}. The significant variability in Fe~II, He~I and O~I emission suggests that the efficiency of the pumping 
channels varies, possibly due to changes in Lyman-series optical depth or other radiative transfer effects. 

\subsection{Location of the emission line formation region}
\label{sec:location}

The transition from P~Cygni to inverse P~Cygni profiles in the Fe~II lines seen in Figure~\ref{fig:FeIIcompare}
is observed in other LBVs, and is interpreted as an indication of infall during the radial contraction phase of 
the stellar photosphere during the \object{S~Dor} variability cycle \citep{wolf90}. In the case of \object{W243},
this transition occurs at a time when the strong Si~II and N~I absorption lines develop broad blue wings\footnote{A similar 
transformation is correlated with periods of enhanced mass-loss in the YHG \object{$\rho$~Cas} \citep{lobel03}.} but 
inspection of the Fe~II multiplet~74 and higher Paschen-series lines suggests that these may not be \textit{true} (inverse) 
P~Cygni profiles, and rather an effect of emission at near-constant radial velocity superimposed on an absorption 
line with a varying line centre: this is similar, but more pronounced, to the effects seen in \object{$\rho$~Cas} 
where static emission lines and variable absorption lines produce apparent line-splitting (see \citealt{lobel98}, figs.~13 and~15). 
With \object{W243}, if the absorption is at its most blueshifted (and especially if a broad blue wing also develops) the 
emission/absorption blend will take on an apparent P~Cygni profile (e.g. as visible in the Pa$\delta$ line in the left 
panel of Figure~\ref{fig:HaPd} in the 10/07/2004, MJD=53192.2 spectrum, or in the Fe~II multiplet~40 lines in 
Figure~\ref{fig:FeIIcompare}), while at its most redshifted the absorption component will be seen entirely on the red 
flank of the emission component, creating an inverse P~Cygni profile: however, neither case corresponds to the significant 
outflow or inflow normally associated with these spectral features. 

The near-static radial velocities of the undistorted emission lines over the course of our observations and the close 
correspondence between the radial velocities of the permitted and forbidden lines suggests that they all form at significant 
radii, when material expanding slowly at constant velocity is ionized by the flux from the hot secondary. As an 
emission spectrum was not observed by \cite{w87}, it appears that this material is a relic of mass-loss during the transition 
from an early-B supergiant to the current cool state. The O~I~$\lambda$8446, Fe~II and higher Paschen-series lines have very 
similar profiles, and all likely form in a common region in the boundary zone between H~I and H~II regions required for efficient 
Ly$\beta$-pumping of the O~I line, while the 
consistency between the measured radial velocities of almost all the permitted and forbidden emission lines 
suggests formation in the same comoving region. The radial velocity of the Ca~II emission is somewhat variable, while
the line is broader than all but the Balmer-series emission lines, most probably reflecting formation in a different 
location: the $\sim$11.9eV ionization potential means recombination will be negligible in a H~II region, and the line
likely results from resonance absorption in an H~I region rather than being limited to the H~I/H~II transition zone. 
It is plausible that the Ca~II emission may be a result of dust evaporation (e.g. \citealt{prieto}) which would indicate 
a post-RSG state for \object{W243}, although the [Ca II] emission that would provide stronger support for this hypothesis 
is absent \citep{shields99}. These lines lie in a region of heavy telluric contamination, but we also note that [Ca~II] is 
only weakly detected in the B[e] star \object{MWC 349A} despite strong Ca~II emission being present \citep{and96}. The 
inferred electron density in the line formation region is high \citep{hs88}, and it is likely that the Ca~II 3$^{2}$D level 
is collisionally de-excited before the forbidden transition can occur. This also appears likely in the case of \object{W243},
where relatively strong [Ca~II] emission would be required to be detectable amidst the telluric lines.

\subsection{Conclusions and future work}

Our observations of \object{W243} show that it has remained in a quiescent, cool-phase state since 2003, with
no indication of either further movement redwards or a return to its former hot state. The $\sim$A3Ia$^{+}$ absorption-line 
spectrum displays photospheric pulsations and episodic mass loss that appears similar to warm YHGs (e.g. \object{$\rho$~Cas}, 
\citealt{lobel98} or \object{W265}, \citealt{ritchie09}). The high Nitrogen abundance and depletion of Oxygen and Carbon imply 
that CNO-processed material is on the surface of \object{W243}, indicating an advanced evolutionary state and significant previous 
mass-loss. Superimposed on the YHG absorption-line spectrum are emission 
lines of H, He~I, Ca~II, Fe~II and O~I formed at large radii when material lost from the LBV is ionized by an unseen hot 
companion. We are unable to strongly constrain the emission-line formation region and the origin of the peculiar, 
double-peaked Balmer-series profiles or the highly variable He~I emission lines. However, high-resolution infra-red
spectroscopy with adaptive optics offers the opportunity to directly probe the line-forming region and thereby determine
the recent mass-loss history of \object{W243}. In addition, abundance studies of the coeval population of both pre- and post-LBV transitional 
stars in Wd1 will allow the evolutionary state of both \object{W243} and its evolutionary contemporaries to be determined. 
 
\begin{acknowledgements}

JSC gratefully acknowledges the support of an RCUK fellowship.  
This research is partially funded by grants AYA2008-06166-C03-02,
AYA2008-06166-C03-03 and Consolider-GTC CSD-2006-00070 from the 
Spanish Ministerio de Ciencia e Innovaci\'on (MICINN). We thank
the referee, Otmar Stahl, for valuable comments.

\end{acknowledgements}

\appendix

\section{List of identified absorption lines in the spectrum of \object{W243}}

\object{W243} displays a large number of absorption lines from neutral and singly-ionized metals; these are listed by species and multiplet 
in Table~\ref{tab:id}. To identify the lines we followed the approach of \cite{oud98}, first identifying strong absorption lines  
and then searching for other lines in the same multiplet using the line identification tables of \cite{moore45}. We then searched for  
other multiplets of the same species, and regard as robust any identification where several lines from the same multiplet are unambiguously 
detected. In a few cases tentative identifications are made when only one line in a given multiplet is detected, provided that one of the following
conditions apply:
\begin{itemize}
\item the multiplet contains no other lines, or the other expected lines in the multiplet are outside the range of coverage of the spectrum.
\item the proposed identification is from a species that has been identified in other multiplets, the line is not close
to possible interstellar or telluric features and no other plausible identification exists. 
\item the proposed identification is from an otherwise unidentified species, but the line is close to a robustly identified line (allowing the 
difference between line centres $\delta\lambda = \lambda_{\text{known}} - \lambda_{\text{unknown}}$ to be measured accurately) and no other 
plausible identification exists with the measured $\delta\lambda$. Nevertheless, such an identification should be considered tentative. 
\end{itemize}

The majority of absorption lines in the range 4800 to $\sim$7000$\text{\AA}$ could be positively 
identified in this manner. After eliminating obvious telluric features, many remaining absorption 
features could be identified as being due to diffuse interstellar bands by comparison with the 
Diffuse Interstellar Band Catalog\footnote{http://leonid.arc.nasa.gov/DIBcatalog.html}, which is 
based on \cite{jd94}. The large number of telluric features above 7000$\text{\AA}$ made 
comprehensive identification in that region problematic, and our list is incomplete in that 
wavelength range with only the stronger features being identified. 

\longtab{1}{
\begin{longtable}{c|c|l}
\caption{\label{tab:id}Identified absorption lines in the spectrum of \object{W243} between 
4800-9000$\text{\AA}$.}\\
        & Multiplet$^{\dag}$ & Rest wavelength  ($\text{\AA}$) \\
\hline
\endfirsthead
\caption{continued.}\\
        & Multiplet$^{\dag}$ & Rest wavelength(s) ($\text{\AA}$) \\
\hline
\endhead
\hline
\endfoot
\hline
Ca~II   & 2  & 8498.02$^{a}$, 8542.09$^{a}$, 8662.14$^{a}$\\
        &uncl.& 8912.07,  8927.36$^{a}$\\
\hline
Cr~II   & 24 & 5153.49, 5305.85 \\
        & 29 & 5369.25, 5395.41, 5409.28, 5430.41\\
        & 30 & 4812.35, 4824.13, 4836.22, 4848.24, 4864.32, 4876.41\\        
        & 43 & 5232.50, 5237.34, 5274.99$^{b}$, 5280.08, 5308.44, 5310.70, 5313.59, 5334.88, 5337.79$^{b}$ \\
        & 50 & 5502.05$^{b,c}$, 5508.60\\
        & 105& 6053.48, 6226.66\\
\hline
Fe~II   & 34 & 6217.95, 6219.54, 6229.34, 6239.36 \\
        & 35 & 5100.55, 5120.34\\
        & 36 & 4993.36, 5036.92\\
        & 40 & 6369.45$^{b}$, 6432.65$^{a}$, 6516.05$^{a}$\\
        & 41 & 5256.89, 5284.09\\ 
        & 42 & 4923.92, 5018.43$^{a}$, 5169.03 \\
        & 46 & 5991.38$^{a}$, 6044.53, 6084.11$^{a}$, 6113.32, 6116.04 \\ 
        & 48 & 5264.80, 5337.71$^{b}$, 5362.86, 5414.09\\
        & 49 & 5197.57$^{a}$, 5234.62, 5254.92, 5275.99$^{b}$, 5316.61$^{a}$, 5325.55, 5346.56, 5425.27, 5477.67\\
        & 55 & 5432.98, 5534.86$^{a}$, 5591.38\\
        & 57 & 5627.49, 5657.92, 5725.95 \\
        & 72 & 7289.05$^{b}$, 7533.42$^{a}$\\
        & 73 & 7222.39$^{a}$, 7307.97$^{a}$, 7320.70$^{a}$, 7462.38$^{a}$, 7515.88$^{a}$, 7711.71$^{a}$\\
        & 74 & 6147.73$^{b}$, 6149.24$^{b}$, 6238.38$^{a}$, 6247.56$^{a}$, 6407.30$^{a}$, 6416.90$^{a}$, 6456.38$^{a}$ \\
        & 163& 6179.38 \\
        & 182& 5952.49\\
        & 199& 6331.97, 6446.43 \\
        & 200& 6103.54, 6175.16, 6305.32 \\        
        &uncl.& 5387.14, 5427.83, 5466.02, 5503.40, 5510.78, 5588.15, 5645.40, 5962.40, 6157.12\\
        &     & 6317.99, 6383.75, 6385.47, 6396.31, 6442.95, 6491.28, 6623.07, 6644.25, 6729.86\\
        &     & 8489.69, 8722.14\\
\hline
H       & 9  & 8598.39 (Pa14), 8665.02 (Pa13), 8750.47 (Pa12), 8862.79$^{a}$(Pa11), 9014.91$^{a}$(Pa10) \\
        & 10 & 8413.32 (Pa19), 8437.96 (Pa18), 8467.25$^{b}$ (Pa17), 8502.49 (Pa16), 8545.38 (Pa15)\\
\hline
Mg~I    & 2  & 5167.32, 5172.68, 5183.60\\
\hline
Mg~II   & 8  & 7877.13, 7896.37$^{f}$ \\
        & 23 & 6545.97$^{c}$\\
\hline
N~I     & 1  & 8680.24$^{b,e}$, 8683.38$^{b}$, 8686.13$^{b}$, 8703.24$^{e}$, 8711.69$^{e}$, 8718.82$^{e}$, 8728.88$^{e}$, 8747.35 \\
        & 2  & 8184.80, 8187.95, 8200.31, 8210.63, 8216.28, 8223.07$^{b}$, 8242.34\\
        & 3  & 7423.63, 7442.28, 7468.29 \\
        & 8  & 8567.74, 8594.01$^{b}$, 8629.24, 8655.88$^{b}$\\
        & 21 & 6481.71$^{b}$, 6482.70$^{b}$, 6483.75$^{b}$, 6484.80$^{b}$, 6506.45 \\
        & 30 & 6758.60 \\
        & 31 & 6706.20, 6723.12, 6733.48 \\
        & 40 & 6752.40 \\  
\hline
O~I     & 1  & 7771.96$^{b}$, 7774.18$^{b}$, 7775.40$^{b}$ \\
        & 4  & 8446.35$^{a}$\\
\hline
S~II    & 6  & 5643.76$^{c}$\\
        & 11 & 5578.85, 5616.63 \\
\hline
Sc~II   & 21 & 5854.31$^{c}$\\ 
        & 28 & 6245.63\\
        & 29 & 5640.97, 5669.03, 5684.19 \\
\hline
Si~II   & 2  & 6347.09$^{e}$, 6371.36$^{b}$\\
        & 4  & 5957.61, 5978.97\\
        & 5  & 5041.06, 5056.02 \\        
        &uncl.& 6660.52, 6665.00, 6671.88, 6699.38\\
\hline
Ti~II   & 69  & 5336.81, 5381.02, 5418.80\\
        & 70  & 5188.70, 5226.53, 5262.10 \\             
        & 86  & 5129.14, 5185.90 \\
        & 92  & 4805.11\\        
        & 103 & 5268.62\\
        & 109 & 5473.52\\ 
        & 112 & 6717.91, 6680.26 \\
        & 114 & 4874.02, 4911.21 \\
        & 298 & 5259.98\\
\hline
V~II    & 55 & 5240.97, 5271.26, 5280.62 \\
        & 97 & 6028.26$^{d}$\\
        &238 & 5642.01$^{c}$\\
        &239 & 5341.22$^{d}$\\
\hline
\hline
\end{longtable}
$^{\dag}$Multiplet from \cite{moore45}. Lines that are unclassified are marked as `uncl.'.\\
$^{a}$inverse P~Cygni profile in 21/09/2003 spectrum\\
$^{b}$blended with nearby line\\
$^{c}$uncertain\\
$^{d}$highly uncertain, very weak\\
$^{e}$excess absorption apparant on blue wing\\
$^{f}$possible P~Cygni profile\\
}

\end{document}